\documentclass[runningheads]{llncs}
\usepackage[T1]{fontenc}
\usepackage{graphicx} %
\usepackage{amsmath,amsfonts,amssymb}
\usepackage{comment}
\usepackage{url}
\usepackage[numbers]{natbib}

\usepackage{algorithm}
\usepackage{algorithmic}

\usepackage{xspace}
\usepackage{xcolor}
\usepackage{dsfont}
\usepackage{booktabs}
\usepackage{placeins}

\newcommand{\FS}{FirstSwap\xspace}
\newcommand{\projname}{FirstSwap\xspace}

\newcommand{\tx}{\ensuremath{\mathsf{tx}}}
\newcommand{\traderequest}{trade\xspace}
\newcommand{\trade}{\ifmmode\tx\else trade\xspace\fi}
\newcommand{\amt}{\mathsf{amt}}

\newcommand{\limit}{\mathsf{limit}}
\newcommand{\price}{\mathsf{price}}
\newcommand{\pricevec}{\mathbf{price}}
\newcommand{\filled}{\mathsf{fill}}
\newcommand{\filledvec}{\mathbf{fill}}
\newcommand{\FR}{\ensuremath{\eta}}
\newcommand{\myapproved}[1]{\ensuremath{\mathsf{A}_{#1}}}
\newcommand{\mycerts}[1]{\ensuremath{\Pi_{#1}}}  

\newcommand{\scadmissible}{\FR{-}admissible\xspace}

\newcommand{\AC}[1]{\ensuremath{\pi(#1)}}  %
\providecommand{\E}{\mathbb{E}}

\providecommand{\eps}{\varepsilon}

\usepackage{needspace}
\usepackage[hidelinks]{hyperref}
\usepackage{cleveref}

\title{\FS: Pre-Consensus Trade Approvals for Decentralized Market Design}
\author{William Diamond\inst{1} \and
Fatima Elsheimy\inst{2} \and
Dahlia Malkhi\inst{3} \and
Geoff Ramseyer\inst{4} \and
Srivatsan Sridhar\inst{5}\thanks{We thank Malcolm Granville for insights in the early stages of this project.}}
\authorrunning{W. Diamond, F. Elsheimy, D. Malkhi, G. Ramseyer, S. Sridhar}
\institute{University of Wisconsin--Madison \and
Yale University \and
University of California, Santa Barbara \and
Stellar \and
University of California, Berkeley}
\date{\today}

\begin{document}

\maketitle

\begin{center}
\small\today
\end{center}

\begin{abstract}
Decentralized exchanges typically give traders no execution guarantee until a block is finalized. In the meantime, external prices move, traders strategically delay submission, and block producers can influence inclusion and ordering. We introduce FirstSwap, a DEX built on a constant-function market maker pricing curve that gives traders an early pre-approval before consensus, guaranteeing inclusion and execution within their submitted price limit, with final ordering determined by limit prices rather than by a block producer.

We demonstrate the benefits of trade pre-approvals in an economic model combining auction theory with CEX-DEX arbitrage. When the CEX price changes gradually, FirstSwap rewards earlier traders with better prices, discouraging strategic delay and yielding higher social welfare and gains to AMM liquidity providers than a batch auction. When the CEX price jumps abruptly, FirstSwap yields higher welfare and smaller AMM losses than fair ordering.

The central challenge is delivering fast trade pre-approvals without consensus. We introduce a decentralized, scaled admissibility protocol under which each validator reserves only a fraction of the available liquidity at every price. A trader obtains approval in a single round of messages with a quorum of validators.
Somewhat surprisingly, despite Byzantine validators who may approve trades arbitrarily,
we show that all quorum-certified trades are jointly admissible and execute within their limits.  Our protocol extends naturally to a setting where buy and sell orders can be pre-approved against each other before a block is finalized.

\end{abstract}

\section{Introduction}
This paper introduces \FS, a decentralized exchange operated by a network of \textit{validators} that finalizes blocks of trades epoch by epoch.
 \FS incentivizes early submission of trade requests by granting pre-approval certificates promptly upon arrival, without waiting for end-of-epoch consensus. Pre-approvals are processed through a single direct interaction between each trader and a quorum of validators, yielding a fast, consensus-free procedure. We develop an economic model to analyze \FS's pre-approval mechanism as a market design. Early pre-approvals yield higher social welfare and gains to AMM liquidity providers than the end-of-block confirmation typical of blockchains. FirstSwap naturally mitigates two prevalent forms of strategic exploitation. First, MEV is curtailed because trade execution depends only on block contents and not on their ordering, and because pre-approvals must come  from a quorum of validators and not at the discretion of a single block builder. Second, CEX--DEX arbitrage exploiting a passive AMM is reduced because traders must race for early pre-approval instead of waiting for arbitrage opportunities. 
 
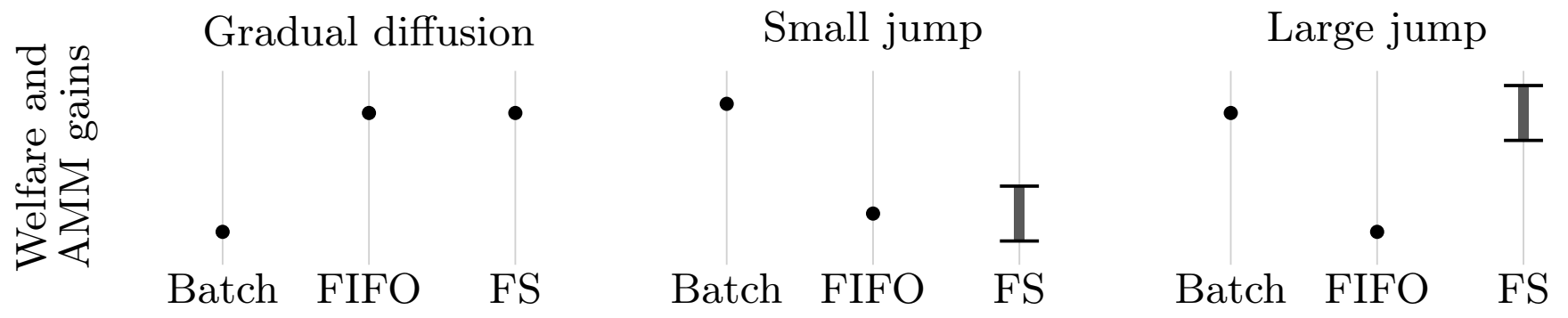
\begin{figure}[htbp]
\vspace{-1em}
\centering
\begingroup
\setlength{\unitlength}{\dimexpr.95\linewidth/480\relax}
\begin{picture}(480,90)
\put(9,45.5625){\makebox(0,0){\rotatebox{90}{\shortstack{Welfare and\\AMM gains}}}}
\put(105,84.375){\makebox(0,0){Gradual diffusion}}
\put(60,12.375){\color{black!18}\linethickness{0.35pt}\line(0,1){59.625}}
\put(60,5.0625){\makebox(0,0){Batch}}
\put(105,12.375){\color{black!18}\linethickness{0.35pt}\line(0,1){59.625}}
\put(105,5.0625){\makebox(0,0){FIFO}}
\put(150,12.375){\color{black!18}\linethickness{0.35pt}\line(0,1){59.625}}
\put(150,5.0625){\makebox(0,0){FS}}
\put(260,84.375){\makebox(0,0){Small jump}}
\put(215,12.375){\color{black!18}\linethickness{0.35pt}\line(0,1){59.625}}
\put(215,5.0625){\makebox(0,0){Batch}}
\put(260,12.375){\color{black!18}\linethickness{0.35pt}\line(0,1){59.625}}
\put(260,5.0625){\makebox(0,0){FIFO}}
\put(305,12.375){\color{black!18}\linethickness{0.35pt}\line(0,1){59.625}}
\put(305,5.0625){\makebox(0,0){FS}}
\put(415,84.375){\makebox(0,0){Large jump}}
\put(370,12.375){\color{black!18}\linethickness{0.35pt}\line(0,1){59.625}}
\put(370,5.0625){\makebox(0,0){Batch}}
\put(415,12.375){\color{black!18}\linethickness{0.35pt}\line(0,1){59.625}}
\put(415,5.0625){\makebox(0,0){FIFO}}
\put(460,12.375){\color{black!18}\linethickness{0.35pt}\line(0,1){59.625}}
\put(460,5.0625){\makebox(0,0){FS}}
\put(60,22.5){\circle*{4.8}}
\put(105,59.0625){\circle*{4.8}}
\put(150,59.0625){\circle*{4.8}}
\put(215,61.875){\circle*{4.8}}
\put(260,28.125){\circle*{4.8}}
\put(370,59.0625){\circle*{4.8}}
\put(415,22.5){\circle*{4.8}}
\put(305,19.6875){\color{black!65}\linethickness{2.2pt}\line(0,1){16.875}}
\put(299,19.6875){\linethickness{0.7pt}\line(1,0){12}}
\put(299,36.5625){\linethickness{0.7pt}\line(1,0){12}}
\put(460,50.625){\color{black!65}\linethickness{2.2pt}\line(0,1){16.875}}
\put(454,50.625){\linethickness{0.7pt}\line(1,0){12}}
\put(454,67.5){\linethickness{0.7pt}\line(1,0){12}}
\end{picture}
\endgroup
\caption{Qualitative comparison of exchange mechanisms (Batch = batch auction, FIFO = first-in-first-out, FS = \projname). Both overall welfare and AMM gains from trade follow the same trends, under the conditions of \Cref{eq:logtwo} and \Cref{lem:jump-volume-welfare-rankings}. Higher positions indicate higher outcomes within each scenario; vertical distances are not to scale. The ranges indicate that the comparison of FS with FIFO (small jump) or Batch (large jump) is ambiguous without further assumptions.}
\label{fig:welfare-amm-comparison}
\vspace{-1em}
\end{figure}

\FS consists of two key  ingredients, each of potential interest on its own:
\textbf{First}, we propose a pre-approval mechanism which constitutes a new form of auction, where both arrival order and price matter. Traders post trade requests with limit prices; each trade request is either pre-approved or rejected based on a price-contingent capacity without being assigned a final position in the block. At epoch end, a final block is produced with an intra-execution ordering, guaranteed to include all pre-approved trade requests settling within their price limits. We develop an economic model to analyze this mechanism and demonstrate that early pre-approvals yield higher social and AMM profitability compared to other proposed AMM mechanisms. \Cref{fig:welfare-amm-comparison} illustrates our findings.

\textbf{Second}, we propose  a fast, consensus-free implementation for a network of validators to collectively operate the pre-approval system. Each pre-approval is obtained through a single direct exchange between the trader and a quorum of validators, with each validator deciding whether to pre-approve a trade request based solely on local information, requiring no coordination among validators. Somewhat surprisingly, the protocol guarantees that at the end of a block epoch, all pre-approvals obtained in this manner can settle within their price limits. Furthermore, Byzantine validators cannot prevent pre-approved trade requests from settling, and if they attempt to hide pre-approved trade requests, not only do they fail but their action is punishable. A straightforward extension of our protocol allows for simultaneous buy and sell orders to be handled simultaneously, where validators can pre-approve additional buys when they have seen sell trade reqests that could execute against them and conversely.  AMM liquidity providers can also deposit and withdraw funds at the end of the block after orders are executed without impacting our analysis.  

\subsection{Overview}
\textbf{The Exchange Design.}
Assume for now that the protocol is run by a single validator.
\FS allocates trades against a pool of fungible assets for sale. The exchange employs a continuous pricing mechanism identical to that of constant-function market makers (CFMMs). The price per unit $p(x)$ increases  as more assets are purchased, where $x$ is the  quantity already bought.

The pre-approval mechanism operates epoch by epoch.
During an epoch, trade requests in the form $\tx=(\amt,\limit)$ arrive and are admitted to \FS promptly upon arrival. A trader's limit price $\limit$ constrains the total quantity that may be consumed before and including their trade.
For each trade request, the pre-approval system determines whether it can be satisfied based on the trades pre-approved so far. If so, unlike existing blockchain-based platforms, the system returns a pre-approval certificate immediately.
Pre-approved trades are then gathered into a finalized block for execution.
\FS guarantees that all pre-approved trade requests can be executed within their price limits, and the final ordering of trades within the finalized block is computed deterministically.

A key notion in the pre-approval mechanism is \textit{admissibility}:
A set $T$ of trades of the form $(\amt, \limit)$ is admissible if for every limit $\ell$ in $T$, the sum total $x$ of amounts in trades whose limit is $\leq \ell$ satisfies $p(x) \leq \ell$.
Note that if a set is admissible, then executing trades in ascending order of limits allows all trade requests to settle within the requested price limits.
An incoming trade request $\tx$ may be pre-approved if adding the trade request to the set $\myapproved{}$ of previously pre-approved trades preserves admissibility; that is, if $\myapproved{} \cup \{\tx\}$ is admissible.

For illustration, consider the special case where all trade requests are for a single unit and the price function is simply $p(x) = x$, so the price per unit equals the number of units already consumed.
The set $\myapproved{} = \{3, 1,3,5\}$ (writing only limits for each trade) is admissible, because for every $\limit = \ell$, the number of trade requests in $\myapproved{}$ with limit at most $\ell$ is $\leq \ell$; indeed, an execution in sorted order $1,3,3,5$ satisfies all price limits.

Now consider an incoming trade request $\tx=4$. The set $\myapproved{} \cup \{\tx\} = \{3, 1,3,5,4\}$ remains admissible, so the new trade request can be pre-approved, and the new execution order is $1,3,3,4,5$.
A different incoming trade request may be $\tx'=2$. The set $\myapproved{} \cup \{\tx'\} = \{3,1,3,5,2\}$ is not admissible: for $\ell=3$, there are $4$ trade requests whose limits are at most $3$, hence $p\bigl(\{\tx \in \myapproved{} \cup \{2\} \mid \tx \leq 3\}\bigr) = 4 > 3$.
Indeed, no sequenced execution of $T \cup \{\tx'\}$ may satisfy all price limits.
Thus, $\tx'$ should be denied pre-approval.

%

\textbf{Economic Incentives and Market Efficiency in \projname.}
A trade in  \projname has a greater chance of executing when it is sent earlier and when it offers a higher limit price, since trades are processed in the order received and more high-limit-price trades can be approved. Traders rationally offer prices strictly below their asset valuations, to ensure a positive gain from an executed trade, but high-valuation traders offer higher prices and also submit trades earlier.  \projname uses both trade time and limit price to ensure that assets go to traders who value them the most. The sorting method in \projname lives between the extremes of two other MEV-resistant AMM mechanisms. A first-in-first-out fair ordering protocol as in \cite{chan2025mechanism} sets a trade's execution price purely based on arrival time, while a batch auction as in \cite{budish2015hft,canidio2023batching} approves trades only based on limit price and ignores arrival time entirely.  \projname, unlike the other two mechanisms, incentivizes traders to bid strictly less than their valuations.  This lack of honest bidding in \projname   actually increases economic efficiency in a model with the realistic possibility of CEX-DEX arbitrage by reducing the volume of arbitrage trades.

We illustrate the economic value of \projname relative to other market designs in an economic model that combines auction theory with dynamic CEX-DEX arbitrage trading. Our model is the first, to our knowledge, that allows us to compare the social welfare of different AMM trading mechanisms where both the limit price and time of a trade are chosen strategically.  This makes our model ideal for examining market design responses to the CEX-DEX arbitrage problem.  By rewarding traders for trading fast instead of waiting for a better trading opportunity, \projname yields higher utilitarian welfare and gains to AMM liquidity providers than an end-of-epoch batch auction.  This is in contrast to \cite{budish2015hft}, which argues in favor of batch auctions in a setting where there is no motive for traders to wait.  While a fair ordering protocol based on \cite{kelkar2020orderfair,chan2025mechanism} provides similar improvements over a batch auction, \projname provides higher welfare and gains to AMM liquidity providers than fair ordering after a large, sudden jump in the CEX price. 
\textbf{Fast, Consensus-free Pre-approvals.}
Granting pre-approvals one by one requires solving the consensus problem repeatedly among the network of validators.
\FS avoids solving consensus for each trade request to speed up the pre-approval step. We develop in \Cref{sec:protocol} a pre-approval protocol consisting of a single exchange between the trader and a quorum of validators that requires no coordination among validators. 

In a nutshell, consensus-free pre-approval works as follows. The trader sends a trade request to the network of validators.
Each validator considers only the trade requests it has pre-approved locally and returns a signed response.
A trade request that has pre-approvals from a quorum of $q$ distinct validators is pre-approval-certified. In order to guarantee that all pre-approval-certified trade requests form an admissible set without relying on consensus, each validator pre-approves only a certain fraction of the allowed capacity at each price. For $n$ validators, $f$ dishonest validators, and quorum size $q=n{-}f$, the pre-approval fraction is $\FR := \frac{q-f}{n-f}$. For the conventional BFT quorum sizes, $n=3f{+}1$, $q =2f{+}1$ and $\FR=\frac{f+1}{2f+1} \approx \frac{1}{2}$; choosing larger quorum systems, e.g., $q=\lceil\frac{4n+1}{5}\rceil$ can yield $\FR \approx \frac{4}{5}$.
Importantly, modified admissibility is checked locally by each validator without any coordination among validators. Nevertheless, we show that all trade requests which are pre-approved individually by validators in this manner can be settled within their price limits.

At the end of each epoch, validators run a single BFT consensus instance, each contributing a \textit{mini-block} of its own pre-approved trades. The consensus output is formed from at least $n{-}f$ mini-blocks and orders them into a final block, guaranteeing inclusion of every trade with a pre-approval certificate.
\section{Fast Pre-Approvals as a Market Design Tool}
\label{sec:auction}

To analyze the economic properties, we first describe the exchange mechanism, as if run by a single validator,
and assume messages are delivered in the order they are sent.
\Cref{thm:fs-validator-count} shows that the same economic allocations would also occur in a range of distributed implementations.
\Cref{sec:protocol} presents one Byzantine fault-tolerant protocol under partial synchrony. 

A \emph{\traderequest} is a tuple
$
    \tx=(\amt,\limit),
$
where $\amt \in \mathbb{R}_{+}$ is the amount of assets the trader wants to buy, and $\limit \in \mathbb{R}_{+}$ is the maximum price per unit.
As in a constant-function market maker (CFMM) \cite{bancor2017},
we consider a
monotonically increasing
price function $p$ such that if $x$ units of assets have already been bought from the exchange,
the price per unit paid for buying the next infinitesimal quantity is $p(x)$.

The exchange operates in periods of time called epochs.
In each epoch, traders send their \traderequest{s} to the exchange.
The exchange approves the \traderequest if adding this \traderequest to the set of all \traderequest{s} approved so far keeps the set \emph{admissible}.

\begin{definition}[Admissibility of \traderequest{s}] \label{def:admissible}
    Let $S=\left\{\tx_i=(\amt_i, \limit_i)\right\}_{i=1}^m$ be a set of \traderequest{s}.
    $S$ is admissible if 
    for all $i$, $p\left(\sum_{j:\limit_j\leq\limit_i} \amt_j\right) \leq \limit_i$.
\end{definition}

Let $S = \left\{\tx_i=(\amt_i,\limit_i)\right\}_{i=1}^{m}$ be the set of approved \trade{s} in the epoch.
The validator executes each trade $\tx_i$ at the price $p\left( \allowdisplaybreaks \sum_{j:\limit_j\leq\limit_i}\amt_j\right)$.
This is equivalent to ordering the approved \trade{s} by their limit price, then executing them at the price determined by the pricing curve.
Admissibility ensures that all trades can be charged prices consistent with the pricing function while satisfying their limit prices. Ordering by limit prices is without loss of generality due to \Cref{lem:order-invariance}.
Thus, while we could choose any such ordering, ordering by limit prices naturally favors trades with higher valuations.

\begin{lemma}
\label{lem:order-invariance}
    Let $S=\left\{\tx_i=(\amt_i, \limit_i)\right\}_{i=1}^m$ be a set of \traderequest{s}.
    There exists an ordering $\left(\tx_i=(\amt_i, \limit_i)\right)_{i=1}^m$ on $S$ such that
    for all $i$, $p\left(\sum_{j \leq i} \amt_j\right) \leq \limit_i$
    if and only if $S$ is admissible.
\end{lemma}

\subsection{Economic Model}

The model has a continuum of infinitesimal traders and time is measured continuously.
At time $t$, trader $i$ has value $\theta_{i,t}=p_{\mathrm{ex},t}+\eps_i$ for holding the asset provided by the AMM, where $p_{\mathrm{ex},t}$ is a publicly observed ``CEX price'' and $\eps_i\sim F_\eps$ is drawn i.i.d. across traders and is fixed over time. The distribution $F_\eps$ has unit mass and is normalized to mean 0, so that $p_{\mathrm{ex},t}$ reflects the valuation of the average trader.  We write $\eps(r)=F_\eps^{-1}(1-r)$ for the value of $\eps$ with a fraction $r$ of the population with a higher valuation  for $r \in (0,1)$.   We make the simple choice that only the CEX price evolves over time, since only arbitrage trades are time-sensitive enough to meaningfully vary in value within a single epoch.

We analyze the \projname protocol first when the CEX price evolves gradually as a diffusion,
\[
    \text{Diffusion model:} \qquad p_{\mathrm{ex},t}=p_{\mathrm{ex},0}+\sigma_{\mathrm{ex}} W_t,
  \qquad
  T_{\mathrm{blk}}\sim\operatorname{Exp}(\lambda_{\mathrm{blk}}),
\]
 where $W_t$ is Brownian motion and the exponentially distributed random time $T_{\mathrm{blk}}$ at which the epoch ends and transactions execute is independent of $W_t$. This Brownian-price/Poisson-epoch timing follows the approach of \cite{milionis2024fees,milionis2022lvr} to tractably study loss-versus-rebalancing against an AMM.  We adapt their approach in a model with rational traders with differing values who can strategically wait to trade. We also examine the case when the CEX price experiences a single large jump $\Delta p_{\mathrm{ex}}$ and $p_{\mathrm{ex},t} = p_{\mathrm{ex},0} + \Delta p_{\mathrm{ex}}, \forall\, t > 0.$

\noindent Economic incentives for traders differ markedly in these settings.  With gradual information, traders will strategically wait to observe the evolution of $p_{\mathrm{ex},t}$ before trading, while they will race each other for good price execution \citep{budish2015hft} after a jump. Each active trader may buy once. Passive LPs commit ex ante to providing AMM liquidity at a price $p(x)$, where $x$ is cumulative inventory sold.

\textbf{Measures of Economic Value}
We compare our trading mechanisms using two criteria: social welfare and AMM gains from trade. In the continuum model, social welfare at the end of the epoch, $T_{\mathrm{blk}}$, is
\begin{equation}
\text{Social welfare:} \qquad
x_0 p_{\mathrm{ex},T_{\mathrm{blk}}}+\int q_i(\eps_i)\eps_i\,dF_\eps(\eps_i),
\end{equation}
where $x_0$ is total asset supply and $q_i(\eps_i)$ is the final quantity allocated to a trader with private value $\eps_i$. For fixed $x_0$, the common-value component $x_0p_{\mathrm{ex},T_{\mathrm{blk}}}$ is identical across allocations. Thus welfare differences come from allocating more of the asset to traders with higher private values. We treat active traders and passive AMM liquidity providers as ex ante drawn from the same population; because passive liquidity must be treated identically by the mechanism, its allocation is uncorrelated with $\eps_i$. Hence, increasing the covariance between active-trader holdings and $\eps_i$ increases overall social welfare.

Our second metric is AMM gains relative to passively holding the asset:
\begin{equation}
\text{AMM gains:} \qquad
(x_0-x_{\mathrm{LP}})
\left(\bar p-p_{\mathrm{ex},T_{\mathrm{blk}}}\right),
\end{equation}
 where $x_{\mathrm{LP}}$ is the inventory remaining with passive liquidity providers and $\bar p$
is the average execution price received by the AMM across trades executed. Because AMM liquidity is passive, strategic traders can leave the AMM with less inventory precisely when $p_{\mathrm{ex},T_{\mathrm{blk}}}$ is high.  This results in ``loss-versus-rebalancing'' or LVR, where the passive AMM liquidity experiences losses after large price movements. An ideal protocol therefore induces traders with high private valuations $\eps_i$ to buy without inducing a large negative correlation between the volume $x_0-x_{\mathrm{LP}}$ executed and price movements $\bar{p}-p_{\mathrm{ex},T_{\mathrm{blk}}}.$

\textbf{The trader's problem with gradual price evolution}
\label{par:auction-diffusion}
A trader chooses both when to submit a trade and the limit price attached to that trade. Let $\tau$ be the stopping time at which the trade occurs and let $p_{\mathrm{bid},\tau}$ denote the limit price submitted at that time. If a mass $x_\tau$ of traders has already been approved to trade, the  marginal AMM price for our trader is $p(x_\tau)$.  Thus a trader with private value $\eps$ has expected utility
\begin{equation}
\text{Utility:} \qquad
\sup_{\tau,p_{\mathrm{bid},\tau}}
\E\left[
\mathbf{1}\{\tau<T_{\mathrm{blk}}\}
\mathbf{1}\{p_{\mathrm{bid},\tau}\geq p(x_\tau)\}
\left(p_{\mathrm{ex},\tau}+\eps-p(x_\tau)\right)
\right].
\label{eq:trader-problem}
\end{equation}
We assume in our gradual diffusion model that messages arrive at validators in exactly the same order as they are sent, so the bid will fill precisely if $p_{\mathrm{bid},\tau} \geq p(x_\tau)$. Since the execution price is $p(x_\tau)$ and does not depend on the trader's bid, whenever the trader wishes to trade it is without loss to submit the smallest limit price that executes,
$
p_{\mathrm{bid},\tau}=p(x_\tau).
$
Thus, the economically substantive decision is when to trade. We look for a monotone equilibrium in which a trader with valuation $\eps(r)$ chooses a time to trade that is strictly increasing in $r$. 
Given the AMM pricing function $p(.)$, a price $p(r)$ is available to a trade that fills after a mass $r$ of trades has previously occurred, so $p_{\mathrm{bid},\tau}=p(r)$ in a monotone equilibrium.  We show in \Cref{app:strategic-timing} that each trader optimally follows a simple ``memoryless'' trading strategy: a trader of type $r$ chooses to submit when the CEX price $p_{\mathrm{ex},t}$ first reaches some threshold $p_{\mathrm{ex}}^{*}(r)$ regardless of other history.

Let
$\gamma_{\mathrm{diff}}\equiv\frac{\sqrt{2\lambda_{\mathrm{blk}}}}{\sigma_{\mathrm{ex}}}.$
The probability that Brownian motion reaches $p_{\mathrm{ex}}^{*}(r)$ before the epoch ends is $e^{-\gamma_{\mathrm{diff}}[p_{\mathrm{ex}}^{*}(r)-p_{\mathrm{ex},0}]}$ (the Brownian first-passage probability formula). Hence the expected utility of a trader whose valuation $\eps(r)$ of the asset is less than that of a fraction $r$ of the population (i.e. $\Pr\{\eps_i > \eps(r)\} = r$) is
\begin{equation}
U(r)=
e^{-\gamma_{\mathrm{diff}}[p_{\mathrm{ex}}^{*}(r)-p_{\mathrm{ex},0}]}
\left[p_{\mathrm{ex}}^{*}(r)+\eps(r)-p(r)\right].
\end{equation}
 A particularly simple equilibrium obtains when
$
\kappa\equiv-\frac{p'(r)}{\eps'(r)}>0$
is constant. In this case, we show in \Cref{app:strategic-timing} that the optimal threshold is
\begin{equation}
p_{\mathrm{ex}}^{*}(r)=p(r)-\eps(r)+\frac{1}{\gamma_{\mathrm{diff}}(1+\kappa)}
. \label{eq:rationalbid}
\end{equation}
Traders optimally wait until their contemporaneous valuation exceeds the AMM price by a constant $\frac{1}{\gamma_{\mathrm{diff}}(1+\kappa)}$.
 For our diffusion analysis, we assume that no trader has a value of $\eps(r)$ so high that $p(r)-\eps(r)$ is below the starting CEX price $p_{\mathrm{ex},0}$, which prevents an initial discrete spike in trading volume.
 
\textbf{Welfare and AMM gains under
\projname and a batch auction.}
Given the rational bidding strategy in \Cref{eq:rationalbid}, type $r$ trades when some value of $p_{\mathrm{ex},t}$ exceeds $p_{\mathrm{ex}}^{*}(r)$ before the epoch ends. This has probability (using the Brownian first-passage probability)
\begin{equation}
\Pr\nolimits_{\mathrm{FS}}^{\mathrm{diff}}(r)
=\exp\left(-\frac{1}{1+\kappa}\right)
\exp\{-\gamma_{\mathrm{diff}}[p(r)-\eps(r)-p_{\mathrm{ex},0}]\}.
\label{eq:auction-diffusion-fill}
\end{equation}
We show in \Cref{app:strategic-timing} that expected welfare $\mathcal W_{\mathrm{FS}}^{\mathrm{diff}}$ and AMM gains $\mathcal R_{\mathrm{FS}}^{\mathrm{diff}}$ are respectively given by
\begin{align}
\mathcal W_{\mathrm{FS}}^{\mathrm{diff}}&=\int_0^1 \eps(r)\Pr\nolimits_{\mathrm{FS}}^{\mathrm{diff}}(r)\,dr,\label{eq:auction-diffusion-welfare}\\
\mathcal R_{\mathrm{FS}}^{\mathrm{diff}}&=\int_0^1\left[\eps(r)-\frac{1}{\gamma_{\mathrm{diff}}(1+\kappa)}\right]\Pr\nolimits_{\mathrm{FS}}^{\mathrm{diff}}(r)\,dr.\label{eq:auction-diffusion-gains}
\end{align}
We compare these metrics to what would happen in a pure batch auction that executes at the end of the epoch.  Here, all bidders will strategically wait until the epoch end and bid based on the final value of $p_{\mathrm{ex},T_{\mathrm{blk}}}$, but (since every trader is infinitely small) they will choose a limit price equal to their true honest values.
 Rank $r$ trades if $p_{\mathrm{ex},T_{\mathrm{blk}}}\ge p(r)-\eps(r)$. Since $p_{\mathrm{ex},T_{\mathrm{blk}}}-p_{\mathrm{ex},0}$ is a Brownian motion at an exponential stopping time, it follows a Laplace distribution, so the probability of this trade is
\begin{equation}
\Pr\nolimits_{\mathrm{batch}}^{\mathrm{diff}}(r)
=\frac12\exp\{-\gamma_{\mathrm{diff}}[p(r)-\eps(r)-p_{\mathrm{ex},0}]\}.
\end{equation}
If $r_{\mathrm{batch}}$ is the rank of the trader choosing a limit price equal to  the batch auction's clearing price, the uniform batch price is
$p_{\mathrm{batch}}^{\mathrm{diff}}=p(r_{\mathrm{batch}})=p_{\mathrm{ex},T_{\mathrm{blk}}}+\eps(r_{\mathrm{batch}})$. Because $p'(r)=-\kappa\eps'(r)$, $p(r)+\kappa\eps(r)$ is constant. Conditional on rank \(r\) executing, the terminal public price exceeds \(p(r)-\eps(r)\) by an exponentially distributed amount with mean $\frac{1}{\gamma_{\mathrm{diff}}}$, so
$\E[p_{\mathrm{batch}}^{\mathrm{diff}}-p_{\mathrm{ex},T_{\mathrm{blk}}}\mid r\text{ executes}]
=\eps(r)-\frac{1}{\gamma_{\mathrm{diff}}(1+\kappa)}.$
 Utilitarian welfare $\mathcal W_{\mathrm{batch}}^{\mathrm{diff}}$ and $AMM$ gains $\mathcal R_{\mathrm{batch}}^{\mathrm{diff}}$ are given by
\begin{equation}
\mathcal W_{\mathrm{FS}}^{\mathrm{diff}}=2e^{-1/(1+\kappa)}\mathcal W_{\mathrm{batch}}^{\mathrm{diff}} \qquad \text{and} \qquad \mathcal R_{\mathrm{FS}}^{\mathrm{diff}}=2e^{-1/(1+\kappa)}\mathcal R_{\mathrm{batch}}^{\mathrm{diff}}.
\label{eq:auction-diffusion-comparison}
\end{equation}

So long as the mean gains from trade are nonnegative for the AMM, \projname dominates a batch auction for both welfare and AMM gains iff
\begin{equation}
\frac{1}{1+\kappa}<\log 2
\quad\Longleftrightarrow\quad
\kappa>\frac{1}{\log 2}-1\approx0.4427.
 \label{eq:logtwo}
\end{equation}
A large value of $\kappa$ is the natural assumption in financial markets, since the common-value variation $p_{\mathrm{ex},t}$ in an asset's value should be much larger than variation in agents' private valuations $\eps$.  We note that a FIFO/fair ordering protocol would yield exactly the same allocation as FS under gradual price diffusion, because the FS equilibrium is in fact providing price execution that is strictly better for earlier trades and no trades are rejected in equilibrium.

\textbf{Optimal Trading Strategies with a Jump.}
Next, we analyze our protocol when the price $p_{\mathrm{ex},t}$ experiences a single large jump instead of gradually changing over time, taking on a single value
\[
p_{\mathrm{ex},\mathrm{jump}}\equiv p_{\mathrm{ex},0}+\Delta p_{\mathrm{ex}}.
\]
In this setting, many traders will submit their orders at once, and we model the arrival times of these orders as uniformly random.  We examine two competitor mechanisms to \projname: the batch auction we examined previously and the incentive-compatible protocol of Chan, Wu, and Shi~\citep{chan2025mechanism} that relies on fair ordering of transactions that we will call FIFO.  In our model, each trader chooses its limit price not knowing the speed with which the trade will be received by validators.  A rational trader maximizes its expected utility across possible realizations where the trade arrives early or late.

 Let $p_{\mathrm{bid},\mathrm{FS}}^{\mathrm{jump}}(r)$ denote the equilibrium bid of a trader with private valuation $\eps(r)$ and $\Pr\nolimits_{\mathrm{fill}}(p_{\mathrm{bid},\mathrm{FS}}^{\mathrm{jump}}(r))$ the probability at equilibrium that the trade is approved.
The optimal bid  $p_{\mathrm{bid},\mathrm{FS}}^{\mathrm{jump}}(r)$ of type $r$ maximizes over the choice of $p_{\mathrm{bid}}$
\begin{equation}[p_{\mathrm{ex},\mathrm{jump}}+\eps(r)-p_{\mathrm{bid}}]\Pr\nolimits_{\mathrm{fill}}(p_{\mathrm{bid}}).
\end{equation}
In a ``regular'' equilibrium where limit prices are strictly decreasing in the rank $r$ and $\Pr_{\mathrm{fill}}(p_{\mathrm{bid}})$ is smooth, the limit for any trader choosing $\Pr_{\mathrm{fill}}(p_{\mathrm{bid}}) \in (0,1)$ satisfies a first-order condition
\begin{equation}
p_{\mathrm{bid},\mathrm{FS}}^{\mathrm{jump}}(r)=p_{\mathrm{ex},\mathrm{jump}}+\eps(r)-\frac{\Pr\nolimits_{\mathrm{fill}}(p_{\mathrm{bid},\mathrm{FS}}^{\mathrm{jump}}(r))}{\Pr\nolimits_{\mathrm{fill}}'(p_{\mathrm{bid},\mathrm{FS}}^{\mathrm{jump}}(r))}.
\label{eq:auction-jump-bid-foc}
\end{equation}
In the \projname equilibrium, every trader's limit price is strictly less than its true valuation of the asset.  Increasing your limit raises the odds of the trade being executed, but a trader with a higher limit pays more if the trade executes.  In a batch auction or FIFO, a small trader's limit price has no impact on the price at which it can trade and is only used to determine if the trader would accept the price it is offered.  Because traders should accept a trade precisely when they pay less than their true valuation, traders rationally choose a limit price equal to their true valuation in both mechanisms. Choosing a limit price equal to the trader's true valuation $p_{\mathrm{bid},\mathrm{batch}}^{\mathrm{jump}}(r)=p_{\mathrm{ex},\mathrm{jump}}+\eps(r)$ is a dominant strategy for both of these mechanisms.

\textbf{Welfare and Revenue}
Next, we compare welfare and AMM gains from trade across our three mechanisms: $\mathsf m \in \{FIFO, FS, batch \text{ }auction\}$.  After a realized jump $\Delta p_{\mathrm{ex}}$, welfare and gains have the following simple expressions, where we denote $\eps_{virt}(r)=\eps(r)+r\eps'(r)$ following Myerson's \cite{myerson1981optimal} similar concept of a virtual value used to study revenue in auctions.

\begin{lemma}[Allocation representation after a jump]
\label{lem:jump-allocation-representation}
 When all traders are rational, welfare and AMM gains from trade are respectively given by
\begin{align}
\mathcal W_{\mathsf m}^{\mathrm{jump}}&=\int_0^1 \eps(r)\Pr\nolimits_{\mathsf m}^{\mathrm{jump}}(r)\,dr; \quad  \mathcal R_{\mathsf m}^{\mathrm{jump}}&=\int_0^1\eps_{virt}(r)\Pr\nolimits_{\mathsf m}^{\mathrm{jump}}(r)\,dr.
\end{align}
 The proof is in \Cref{app:jump-allocation}.
\end{lemma}
We use this lemma to compare welfare and revenue across our three mechanisms.  Under conditions we state in \Cref{lem:single-crossing-welfare}, the equilibria induced by these mechanisms  can be ordered with a ``single crossing condition.''  When this condition holds, there is a cutoff $r^*$ where one mechanism has a higher probability of trade execution for $r>r^*$ and the other for $r<r^*$.
\begin{lemma}[Single-crossing welfare and revenue decompositions]
\label{lem:single-crossing-welfare}
Consider two mechanisms $\mathsf m,\mathsf n\in\{\mathrm{FS},\mathrm{FIFO},\mathrm{batch}\}$, with a difference in trade probabilities $d_{mn}(r)=\Pr\nolimits_{\mathsf m}^{\mathrm{jump}}(r)-\Pr\nolimits_{\mathsf n}^{\mathrm{jump}}(r)$ and volumes $\Delta Vol_{\mathsf m \mathsf n}^{\mathrm{jump}}=\int_0^1(\Pr\nolimits_{\mathsf m}^{\mathrm{jump}}(r)-\Pr\nolimits_{\mathsf n}^{\mathrm{jump}}(r))\,dr$.
If  the two mechanisms satisfy the  single-crossing condition with crossing point $r^*$, welfare and AMM gains satisfy
\begin{align}
\label{eq:auction-gains-decomposition}
\mathcal W_{\mathsf m}^{\mathrm{jump}}-\mathcal W_{\mathsf n}^{\mathrm{jump}}
&=\underbrace{\int_0^1[\eps(r)-\eps(r^*)]d_{mn}(r)\,dr}_{\text{selection effect}}
+\underbrace{\eps(r^*)\Delta Vol_{\mathsf m \mathsf n}^{\mathrm{jump}}}_{\text{volume effect}},\\
\mathcal R_{\mathsf m}^{\mathrm{jump}}-\mathcal R_{\mathsf n}^{\mathrm{jump}}
&=\underbrace{\int_0^1[(\eps_{virt}(r)-\eps_{virt}(r^*)]d_{mn}(r)\,dr}_{\text{selection effect}}
+\underbrace{\eps_{virt}(r^*)\Delta Vol_{\mathsf m \mathsf n}^{\mathrm{jump}}}_{\text{volume effect}}. \nonumber
\end{align}

The batch equilibrium always has the single crossing property relative to FIFO or \projname.  Sufficient conditions for \projname to have the single-crossing property relative to FIFO are: 1. $F_\eps$ has a smooth nondecreasing hazard rate $f_\eps(e)/(1-F_\eps(e))$; and
2. $p$ is $C^3$ with $p',p''>0$,  and $1\le \mathcal C(x):=\frac{p'(x)p'''(x)}{p''(x)^2}\le2$ is a nondecreasing function.

\end{lemma}
In the above decomposition, the ``selection effect'' term always yields higher welfare and AMM gains for a batch auction than for \projname or FIFO, since a batch auction perfectly allocates the asset to agents with the highest $\eps$ values.  Because the \projname equilibrium features limit prices strictly below agents' true valuations, it increases the probability that a given price will result in an executed trade.  Under the conditions in the lemma, this results in a positive selection effect for \projname relative to FIFO, since high-valuation agents will rationally choose limit prices at which they pay to increase their chances of trading.

%

%

\begin{lemma}[Jump size, volume, and welfare rankings]
\label{lem:jump-volume-welfare-rankings}

\noindent (i) Let $\Delta p_{\mathrm{mid}}=p_{\mathrm{ex},\mathrm{jump}}-p(1/2)$.
\noindent \textit For any $\Delta p_{mid}$,
$Vol_{\mathrm{FIFO}}^{\mathrm{jump}}\ge\max\{Vol_{\mathrm{batch}}^{\mathrm{jump}},Vol_{\mathrm{FS}}^{\mathrm{jump}}\}.$
Batch's single crossing with either mechanism is at
$\eps^*=\eps(Vol_{\mathrm{batch}}^{\mathrm{jump}})=p(Vol_{\mathrm{batch}}^{\mathrm{jump}})-p_{\mathrm{ex},\mathrm{jump}}$.  Batch  has higher welfare and AMM gains than FIFO for $\Delta p_{\mathrm{mid}}\geq 0$.

\noindent \textit{(ii) Small jumps.}
Under additional  regularity conditions in \Cref{app:batch-sting-jump}, if
$0\le \Delta p_{\mathrm{mid}}<(\frac{1}{2f_\eps(0)}+p'(1/2)/4)\sqrt{[p(1/2)-p(1/4)]/(24p'(1/2))}$, and $p(1/2)-p(1/4)\ge2F_\eps^{-1}(3/4)$ then batch has 
higher welfare and AMM gains than \projname.

\noindent \textit{(iii) Large jumps.}
 If
$p_{\mathrm{ex},\mathrm{jump}}\ge p(1)+\frac12p'(1)$, then \projname has higher welfare and AMM gains than FIFO. The proofs are in \Cref{app:convex-amm-crossing,app:batch-sting-jump}.
\end{lemma}

\textbf{Equivalent allocations with many validators.}
Our economic analysis has assumed that orders submitted at the same time are received by a single validator in random order.  We show in the appendix that the equilibrium of any of our protocols is equivalent in terms of the probability of trade, welfare and AMM gains to one with many validators if every validator's view is used in selecting which transactions to execute at the end of the epoch.  For FS, our equivalence result applies to any version of our protocol satisfying three conditions:
(i) accepted orders execute in ascending bid order at marginal AMM
prices, without exceeding their submitted limits;
(ii) against the equilibrium bids of other traders, bids arbitrarily
close to $p(0)$ from above have positive execution probability; and
(iii) some finite bid guarantees execution.  For FS, this result is akin to the revenue equivalence theorem of Myerson \cite{myerson1981optimal} for standard auctions, showing that within a certain family of mechanisms implementation details do not impact the economic allocation of resources.  The proof is in \Cref{app:validator-count}.

\begin{theorem}[Validator-count equivalence]
\label{thm:fs-validator-count}
For each of the mechanisms batch, FIFO, and \projname, any regular monotone equilibrium in a protocol with multiple validators yields the same allocation,
welfare, and AMM revenue as in the one-validator case.  For batch and FIFO, the bidding strategies are also identical, but the optimal bidding strategies for \projname can vary with the number of validators despite yielding the same allocation. 
\end{theorem}

\section{The \FS Decentralized Protocol}
\label{sec:protocol}

In this section, we describe the decentralized \FS protocol, a distributed implementation of the abstract \FS design presented in~\Cref{sec:auction} over a network of $n$ validators. Due to space limitations, a glossary, algorithms in pseudocode and several (easy) proofs are deferred to~\Cref{app:algorithms}.\\

\noindent \textbf{The fault model.} 
We assume at most $f\leq\lfloor(n-1)/3\rfloor$ validators are
Byzantine; typically $f=\lfloor(n-1)/3\rfloor$, though other thresholds may be used. The numbers of all traders and Byzantine traders are
unknown. Honest parties follow the protocol while Byzantine
parties may deviate arbitrarily. We assume unforgeable digital signatures
and work in the standard partially synchronous model~\cite{dwork1988consensus}:
there exists an unknown Global Stabilization Time (GST) and a known bound
$\Delta_{\text{net}}>0$ such that after GST any message sent at time $t$ is
received by $\max(t,\text{GST})+\Delta_{\text{net}}$. Safety holds at all
times; liveness is guaranteed only after GST.

\textbf{Protocol structure.} The protocol operates in fixed-duration periods called \emph{epochs}; unless
noted otherwise, we focus on a single epoch. 
Within each epoch the protocol
proceeds in two stages: a consensus-free \emph{pre-approval} stage, followed
by a finalization stage that runs a single BFT consensus instance.
For liveness, we assume the epoch is long enough for traders to submit trade
requests and receive validator responses, followed by a period sufficient for one consensus instance.

\textbf{Scaled admissibility.} The key technical challenge is ensuring that all independently pre-approved
requests are jointly admissible, without any coordination
among validators. To this end, each validator pre-approves only a scaled fraction $\FR$
of the available capacity at each price level. For $n$ validators, up to $f$
dishonest, and quorum size $q:=n{-}f$, define the approval fraction

$$\FR:=\frac{q-f}{n-f}~.$$ 

For conventional BFT parameters ($n=3f{+}1$,
$q=2f{+}1$), $\FR=\frac{f+1}{2f+1}\approx\frac{1}{2}$; larger quorums
(e.g.\ $q=\lceil\frac{4n+1}{5}\rceil$) yield $\FR\approx\frac{4}{5}$. 
Recall that a set  $S$ of trade requests is admissible if they can all be satisfied simultaneously within their price limits (\Cref{def:admissible}).
A scaled \scadmissible criterion is defined as follows:

\begin{definition}
[\scadmissible] 
\label{def:fr-admissible}
    The set $S = \{(\amt_i, \limit_i)\}$ is \emph{$\eta$-admissible} if the scaled set $\{(\amt_i/\eta, \limit_i)\}$ is admissible. 
\end{definition}

Since $\eta < 1$, this is a strictly stronger requirement than admissibility, and the slack it introduces is precisely what allows each validator to enforce admissibility locally and independently, without coordinating with other validators, while still guaranteeing that all trades with approval certificates are jointly admissible.

\textbf{Pre-Approval Protocol.}
\label{sec:pre-approval}
During the pre-approval stage, traders submit trade requests
$\tx=(\amt,\limit)$ to each validator and collect signed responses.
A \emph{pre-approval} $\sigma_v(\tx)$ from validator $v$ is a signed message
$\langle\mathsf{APPROVE},\tx\rangle$.
A collection of pre-approvals for $\tx$ from $q$ distinct validators
constitutes an \emph{approval certificate} $\AC{\tx}$, where $q$ is a
Byzantine quorum threshold chosen so that any two quorums intersect in at
least one honest validator (e.g.\ $q=n{-}f$ in standard BFT settings).
A trader that obtains $\AC{\tx}$ may forward it to validators. Each validator $v$ maintains a local set $\myapproved{v}$ of trade requests
it has pre-approved and a set $\mycerts{v}$ of approval certificates it has
 seen. Both sets are empty at the start of an epoch.
A validator stores its first response to each $\tx$ and replays it on
retransmission. Validator $v$ pre-approves $\tx$ if and only if
$\myapproved{v}\cup\{\tx\}$ is \FR-admissible (\Cref{def:fr-admissible}).
If so, $v$ returns $\sigma_v(\tx)$ and updates
$\myapproved{v}:=\myapproved{v}\cup\{(\tx,\sigma_v(\tx))\}$.
Scaled admissibility is checked locally by each validator without any coordination between validators. All trade requests holding approval
certificates, each admitted individually by validators in the above decentralized manner, can be satisfied within
their price limits, without relying on consensus.

\textbf{Finalization Protocol.}
\label{sec:finalization}
Throughout the epoch, traders and validators forward signed pre-approvals and
approval certificates they observe to all validators. At the end of the epoch, each validator $v$ forms and signs a \emph{mini-block} $B_v=(\myapproved{v},\mycerts{v})$.  An honest validator includes in \myapproved{v}
all pre-approvals it signed before the finalization cutoff. 
It also includes in $\mycerts{v}$ every valid certificate received before this cutoff. Anyone can verify the validator did not surpass its scaled pre-approval capacity. More formally: 

\begin{definition}[Well-formed mini-block] \label{def:well-formed}    
A mini-block is \emph{well-formed} if: (i) $\{\tx:\tx\in\myapproved{v}\}$ is
\FR-admissible and each pre-approval is correctly signed by $v$, and (ii)
every $\AC{\tx}\in\mycerts{v}$ is correctly formed. 
\end{definition}

Each validator submits its signed mini-block as input to a single MVBA
consensus instance. The MVBA protocol must satisfy an \emph{external validity} condition such that a decided value $B$ satisfies the predicate $\mathsf{Valid}$ below:

\begin{definition}[Valid]\label{def:valid}
$\mathsf{Valid}(\mathcal{B})$ holds if (i) $\mathcal{B}$ contains well-formed
mini-blocks from at least $M_{B} \geq n{-}f$ distinct validators, and (ii) for every
$\AC{\tx}$ appearing in $\mathcal{B}$ and every validator $v$ whose
mini-block $B_v=(\myapproved{v},\ast)$ is in $\mathcal{B}$, if $v$'s
signature appears in $\AC{\tx}$ then the same signed pre-approval for $\tx$
must appear in $\myapproved{v}$.
\end{definition}

External validity provides three assurances: each $\myapproved{v}$ must
satisfy \FR-admissibility to be counted; a trade request pre-approved by
$n{-}f$ honest validators appears in at least $M_B{-}f$ included mini-blocks;
and a validator that attempts to hide a signed pre-approval it issued may be
excluded from $\mathcal{B}$ via the cross-check.

In addition to external validity, the MVBA must uphold standard consensus properties:
\emph{agreement} (no two honest validators decide different values), 
and \emph{termination} (every honest validator eventually decides after GST,
provided all honest validators submit a valid input). Beyond these
properties, the MVBA protocol is treated as a black box. Once the MVBA decides $\mathcal{B}$, an ordered set of trades is induced
deterministically:

\begin{description}
    \item[Pre-approved inclusion.]
    Every trade request $\tx$ with signatures from at least 
    $\FR M_{B}$
    distinct
    validators in $\mathcal{B}$ is included.

    \item[AC inclusion.]
    Every trade request $\tx$ for which a valid $\AC{\tx}$ appears in
    $\mathcal{B}$ is included, unless already covered above.

    \item[Execution ordering.]
    All included trade requests are ordered in non-decreasing order of
    $\limit$, with ties broken deterministically by trade request identifier.
    Every included trade is guaranteed to be satisfiable within its price
    limit.
\end{description}

\FloatBarrier

\noindent 
\textbf{Security Analysis.}
\label{sec:security}
 \FS guarantees the following desired properties:

\begin{description} \label{def:desired}
    \item[$f$-Agreement.] If at most $f$ validators are Byzantine, all honest validators output the
same final block at each epoch. (Proof deferred to~\Cref{app:algorithms}.)
    \item[$f$-Price validity.] Suppose at most $f$ validators are Byzantine. Let $T$ be the ordered set of
trades in the final block. Then $T$ is admissible (\Cref{def:admissible}).

\item[$(f,\FR)$-Pre-approval liveness.]
    Let $S_{\text{any}}$ be the set of trade requests sent to any honest
    validator during the epoch, and $S_{\text{all}}\subseteq S_{\text{any}}$
    the set sent to all honest validators.
    If $S_{\text{any}}$ is \FR-admissible, then under synchrony every trade request in
    $S_{\text{all}}$ obtains an approval certificate.

\item[$f$-Pre-approval inclusion.]
    If either (i) a trade request $\tx$ has pre-approvals from $q$ honest
    validators, or (ii) an $\AC{\tx}$ is received by $f{+}1$ honest
    validators, then $\tx$ must be included in the final block.

\item[Accountable inclusion.]
There exists a function $J$ such that if $\tx$ obtains an approval
certificate and is not included in the epoch's final block, then there
exists at least one validator $v$ and evidence $w$ with $v\in J(w)$.
Moreover, for all honest validators $v$ and all $w$, $v\notin J(w)$. (Proof deferred to~\Cref{app:algorithms}.)
\end{description}

\noindent
\textbf{Validity.}
Let $V^*$ be the set of creators of the mini-blocks in the decided proposal
$\mathcal B$, and let $M_{B} := |V^*|$. External validity gives $M_{B} \geq n-f$ and
ensures that every $B_v=(\myapproved{v},\mycerts{v})$ for $v\in V^*$ is
well-formed. Let $S$ be the set of trade requests included by the two
inclusion rules of the protocol.
We use the notation $\limit(\tx)$ and $\amt(\tx)$ to denote the limit and amount of a trade request $\tx$.

\begin{lemma}[Representation of included requests]
\label{lem:representation}
Every trade request in $S$ appears in $\myapproved{v}$ for at least 
$\FR M_{B}$
validators $v\in V^*$.
\end{lemma}

\begin{proof}
If a request is included by the pre-approved inclusion rule, the claim follows
from its threshold $\FR M_B$. Otherwise, its approval certificate has $q$
signers, at least $q+M_B-n$ of whom belong to $V^*$. Dividing this lower
bound by the number of selected validators gives
\[
 \frac{q+M_B-n}{M_B}
 =1-\frac{n-q}{M_B}
 \geq 1-\frac{n-q}{n-f}
 =\frac{q-f}{n-f}
 \geq\FR.
\]
The first inequality uses $n-q\geq0$ and $M_B\geq n-f$; the last uses the
assumed bound on $\FR$. Thus at least $\FR M_B$ selected validators signed
the request. The external-validity cross-check requires its signed approval
to appear in $\myapproved{v}$ for every such signer.
\qed
\end{proof}

\begin{lemma}
[Admissibility of included requests]
\label{lem:included-admissible}
The set $S$ is admissible.
\end{lemma}

\begin{proof}
Fix a limit $\ell$ occurring in $S$ and write
$X(\ell)=\sum_{\tx\in S:\limit(\tx)\leq\ell}\amt(\tx)$. For each
$v\in V^*$, similarly write
$X_v(\ell)=\sum_{\tx\in\myapproved{v}:\limit(\tx)\leq\ell}\amt(\tx)$.
By \Cref{lem:representation}, each request counted in $X(\ell)$ appears in
at least $\FR M_B$ selected approved sets. Counting its amount in each gives
\begin{equation}
 \FR M_B X(\ell)\leq \sum_{v\in V^*}X_v(\ell).
 \label{eq:representation-count}
\end{equation}
For every $v$ with $X_v(\ell) > 0$, let $\ell_v$ be its largest
limit at most $\ell$ among the requests counted by $X_v(\ell)$. Since $\myapproved{v}$ is $\FR$-admissible,
$p(X_v(\ell)/\FR)\leq\ell_v\leq\ell$. Let
$Y=\max_{v\in V^*}X_v(\ell)/\FR$, treating an empty prefix as zero. Then
$p(Y)\leq\ell$. Dividing \Cref{eq:representation-count} by $\FR M_B>0$ gives
\begin{equation}
 X(\ell)\leq \frac{1}{\FR M_B}\sum_{v\in V^*}X_v(\ell)
 \leq \frac{1}{\FR M_B}\sum_{v\in V^*}\FR Y=Y.
\end{equation}
Monotonicity of $p$ gives $p(X(\ell))\leq p(Y)\leq\ell$. This holds for
every limit in $S$, which is exactly admissibility.
\qed
\end{proof}

\begin{corollary}
\label{thm:price-validity}
The protocol satisfies $f$-Price validity.
\end{corollary}

\noindent
\textbf{Liveness.}
The following proofs use the synchrony and epoch-duration assumptions from
above. In particular, messages sent during the reservation period
arrive before its cutoff, and MVBA termination produces a final block.

\begin{theorem}
\label{thm:fast-reservation}
The protocol satisfies $(f,\FR)$-pre-approval liveness.
\end{theorem}

\begin{proof}
Assume $S_{\mathrm{any}}$ is $\FR$-admissible. Immediately before an honest
validator $v$ processes $\tx\in S_{\mathrm{any}}$, the set
$\myapproved{v}\cup\{\tx\}$ is a subset of $S_{\mathrm{any}}$. Removing
requests only decreases every prefix amount, so this subset remains
$\FR$-admissible. Validator $v$ therefore signs an approval for $\tx$. Every request in $S_{\mathrm{all}}$ is approved by every honest validator.
There are at least $n-f=q$ honest validators, and synchrony delivers their
responses during the epoch. Thus the trader obtains an AC.
\qed
\end{proof}

\begin{theorem}
\label{thm:reservation-inclusion}
The protocol satisfies $f$-pre-approval inclusion.
\end{theorem}

\begin{proof}$n \geq 4f+1$, $q = n-f$, and $\FR = \frac{q-2f}{n-f}$,
First suppose $\tx$ has approvals from $q$ honest validators. Exactly $n-M_B$
validators lie outside $V^*$, so at least $q+M_B-n\geq\FR M_B$ of these
honest approvers belong to $V^*$, by the bound in the proof of
\Cref{lem:representation}. Their complete approved sets contain their signed
approvals for $\tx$, meeting the pre-approved inclusion threshold.
Now suppose an AC for $\tx$ is received before the cutoff by $f+1$ honest
validators. Since $n-M_B\leq f$, at least one of these validators belongs to
$V^*$. The honest mini-block construction places the AC in that validator's
$\mycerts{v}$, so the AC inclusion rule includes $\tx$.
\qed
\end{proof}

\begin{corollary}
    \label{cor:security}
    For any $n,f,q,\FR$ such that $n \geq 3f+1$, $q = n-f$ and $\FR \leq \frac{q-f}{n-f}$, the FirstSwap protocol with parameters $n,f,q,\FR$ satisfies $f$-agreement, $f$-price validity, $(f,\FR)$-pre-approval liveness, $f$-pre-approval inclusion, and accountable inclusion.
\end{corollary}

\noindent\textbf{Stronger inclusion guarantees for pre-approved trades.}
By parameterizing the protocol with $n = 4f+1$, $q = n-f$, and $\FR = \frac{q-2f}{n-f} = \frac{f+1}{3f+1} \approx \frac{1}{3}$, we can strengthen pre-approval inclusion: if a trade request $\tx$ is pre-approved by $q$ validators (not necessarily all honest), then $\tx$ is guaranteed to be included in the final block.
Moreover, with these parameters, accountable inclusion is also strengthened: at least $f+1$ validators (not just one) is caught with evidence if a pre-approved trade is not included in the final block. 
The proofs of these strengthened properties are in \Cref{app:algorithms}.
All other security properties are preserved because these parameters satisfy the conditions of \Cref{cor:security}.

\section{Maximal Fill Volume}
\label{sec:partial_fills_buy}

The pre-approval liveness guarantee of the \projname protocol in \Cref{sec:protocol} requires the set of trade requests $S_{\text{any}}$ sent to any honest validator to be $\FR$-admissible.
For example, let $p(x)=x$ and suppose two traders submit requests $\tx_1 = (\amt_1=1, \limit_1=5)$ and $\tx_2 = (\amt_2=2, \limit_2=5)$ to all honest validators. For $\FR=1/2$, the set $S_{\text{any}} = \{\tx_1, \tx_2\}$ is \emph{not} $\FR$-admissible, so the liveness guarantee does not apply. No trade is guaranteed to be included. Yet each request is individually $\FR$-admissible, as is the full amount of $\tx_1$ combined with half the amount of $\tx_2$, suggesting that the protocol leaves some liveness unexploited.
Moreover, a malicious trader can spam the system by submitting a large request to just one honest validator. This makes $S_{\text{any}}$ non-$\FR$-admissible, leaving the pre-approval liveness property unable to guarantee inclusion of either the honest requests or the spammer's request.
In contrast, a standard blockchain-based DEX would include either the honest requests or the spammer's request in a block: if the spammer displaced the honest requests, it would at least pay fees for its own.
If no trade requests are pre-approved, no trade requests are guaranteed to be included in the block.

We propose a simple extension  that addresses these limitations by allowing partial fills in the final block.
A \emph{trade} is denoted by $\trade = (\tx, \filled)$, where $\tx=(\amt, \limit)$ is the original \emph{trade request} and $\filled \in [0, \amt]$ is the amount filled. The trader specifies the maximum amount $\amt$ he is willing to buy, and the protocol determines the actual amount bought $\filled$.
Validators continue to use the $\FR$-admissibility condition to pre-approve trade requests, so pre-approval liveness remains as it is.
The final block fully fills all pre-approved requests (preserving pre-approval inclusion) and partially fills residual requests while maintaining admissibility (improving livness). 
The residual fill algorithm solves a fractional knapsack problem variant, maximizing total filled volume.

With partial fills, a set of trades is considered admissible based on their filled amounts $\filled$ rather than their requested amounts $\amt$.
$\FR$-admissibility continues to use the full amounts.

\begin{definition}[Admissible trades]
\label{def:partial-admissible}
Let $T=\{(\tx_i=(\amt_i,\limit_i),\filled_i)\}_{i=1}^m$.
The set $T$ is \emph{admissible} if $0\leq\filled_i\leq\amt_i$ for every $i$
and, for every $i$ with $\filled_i>0$,
$
 p\left(\sum_{j:\limit_j\leq\limit_i}\filled_j\right)\leq\limit_i.
$
\end{definition}

We keep the pre-approval rule and the consensus validity conditions of
\Cref{sec:protocol}. To make rejected requests available for residual
inclusion, each honest validator $v$ also records every valid request it
receives before the finalization cutoff in a set $\mathcal R_v$, and includes
this set in its signed mini-block
$B_v=(\myapproved{v},\mycerts{v},\mathcal R_v)$.
Well-formedness checks the validity and trader authorization of these
additional requests. Recording does not reserve capacity: the
$\FR$-admissibility check still applies only to $\myapproved{v}$.

Let $\mathcal B$ be the collection of mini-blocks decided by consensus,
$V^*$ their distinct creators, and $M_B=|V^*|\geq n-f$.
Let $U$ contain every distinct request appearing in $\mathcal B$,
including recorded but unapproved requests. Write $G$ for
the requests selected by either of the first two inclusion rules below.
As in the main protocol, use $q=n-f$ and $\FR=(n-2f)/(n-f)$.
The final ordered block is constructed as follows.

\begin{description}
    \item[Pre-approved inclusion.] 
    Every trade request $\tx = (\amt, \limit)$ with signatures from at least $M_{B}{-}f$ distinct
    validators in $\mathcal{B}$ is included with $\filled = \amt$.

    \item[AC inclusion.]
    Every trade request $\tx = (\amt, \limit)$ for which a valid $\AC{\tx}$ appears in
    $\mathcal{B}$ is included with $\filled = \amt$, unless already covered above.

    \item[Residual inclusion.]
    Then consider every remaining trade request in $U$ in a fixed deterministic order, and assign it the maximum fill amount, up to its requested amount, that preserves admissibility.
    Initially set the fills of all other requests to zero. At each step keep
    all other fills fixed, and check admissibility against every prefix in
    $U$, including prefixes ending at requests that currently have zero fill.
    Equivalently, assign the largest $x\in[0,\amt_i]$ for which
    replacing $\filled_i=0$ by $x$ keeps the complete fill vector admissible.

    \item[Execution ordering.]
    Discard requests with zero fill and order all remaining requests in
    non-decreasing order of $\limit$, breaking ties deterministically by request identifier. 
\end{description}

\paragraph{Feasible fills and capacity.}
For a set of requests $U$, write
\begin{equation}
 \mathcal F(U)=\left\{\filledvec\in\prod_{i=1}^{|U|}[0,\amt_i]:
 \{(\tx_i,\filled_i)\}_{\tx_i\in U}\text{ is admissible}\right\}.
\label{eq:partial-fill-feasibility}
\end{equation}
Define
\[
 R(U)=\max_{\filledvec\in\mathcal F(U)}\sum_i\filled_i,
 \qquad R(\emptyset)=0.
\]

\noindent\textbf{Security analysis.}
The strengthened liveness property achieved by this modified protocol is stated in \Cref{def:maximal-volume} and proved in \Cref{thm:maximal-volume}. It guarantees that the total filled volume of the final block is at least as large as the maximum feasible fill of all valid trade requests sent to all honest validators before the finalization cutoff.
This holds even if the set of requests sent to honest validators is not $\FR$-admissible. In particular, even if a malicious trader spams some honest validators, either the honest trade requests or an equivalent volume of the spammer's trade requests will be filled in the final block.
This restores a liveness guarantee similar to that of a standard blockchain-based DEX, in which the spammer would pay fees for the spammer's trade requests.

Price validity is preserved by construction: By \Cref{lem:included-admissible}, $G$ is admissible when
$\filled=\amt$ for every request. Assigning zero fills to the remaining
requests changes no filled amount and preserves admissibility. Each residual
step preserves admissibility by construction and leaves pre-approved fills
unchanged. Hence the complete final fill vector is admissible.
All other security properties of the original protocol remain unchanged.

\begin{definition}[$f$-Maximal volume filled]
\label{def:maximal-volume}
Suppose that the network is synchronous and at most $f$ validators are
Byzantine during an epoch. Let $S$ be the set of valid trade requests sent
to all honest validators in time to arrive before the finalization cutoff.
Let $T$ denote the trades in the final block of the epoch. The protocol
satisfies $f$-Maximal volume filled if
\begin{equation}
 \sum_{(\tx_i,\filled_i)\in T}\filled_i\geq R(S).
\end{equation}
\end{definition}

\begin{lemma}[Maximal fills maximize volume]
\label{lem:maximal-volume}
Starting from the full fills of $G$, the protocol's residual
rule produces a feasible fill of $U$ with total volume $R(U)$.
\end{lemma}

\begin{proof}
By \Cref{lem:included-admissible}, the initial full fills of
$G$ are admissible; zero fills for the remaining requests and
each subsequent residual step preserve admissibility.
Moreover, no final fill can be increased while the other fills are fixed.
Pre-approved requests are already full. For a residual request, any increase
that were admissible at the end would also have been admissible when it was
processed: all other fills were then no larger, and decreasing fills
preserves admissibility. Such an increase would contradict the greedy rule.

Suppose, for a contradiction, that another admissible fill vector
$\widehat{\filledvec}$ has larger total volume. Write
$
 D=\sum_j(\widehat{\filled}_j-\filled_j)>0,
$
and choose a request $\tx_i$ with largest limit among those for which
$\widehat{\filled}_i>\filled_i$. For every $\ell\geq\limit_i$, requests
with limit above $\ell$ have $\widehat{\filled}_j\leq\filled_j$, so
\begin{equation}
 \sum_{j:\limit_j\leq\ell}
 (\widehat{\filled}_j-\filled_j)
 =D-\sum_{j:\limit_j>\ell}(\widehat{\filled}_j-\filled_j)
 \geq D.
 \label{eq:maximal-volume-bound}
\end{equation}
Increase $\filled_i$ by
$\varepsilon=\min\{D,\widehat{\filled}_i-\filled_i\}>0$.
The increased fill is at most $\widehat{\filled}_i\leq\amt_i$.
Prefixes below $\limit_i$ are unchanged, and
\eqref{eq:maximal-volume-bound} shows that every prefix at or above
$\limit_i$ has total fill at most the corresponding prefix of
$\widehat{\filledvec}$.

For each such limit $\ell$, the alternative prefix contains a positive
fill because $\widehat{\filled}_i>0$. Let $\ell'$ be the largest limit of
any positive-filled request in that prefix. Admissibility of
$\widehat{\filledvec}$ gives
\[
 p\left(\sum_{j:\limit_j\leq\ell}\widehat{\filled}_j\right)
 =p\left(\sum_{j:\limit_j\leq\ell'}\widehat{\filled}_j\right)
 \leq\ell'\leq\ell.
\]
By monotonicity of $p$, the increased fill vector therefore remains
admissible, including at $\limit_i$. This contradicts the impossibility of
increasing any final fill. No admissible vector has larger total volume,
so the protocol attains $R(U)$.
\qed
\end{proof}

\begin{theorem}[$f$-Maximal volume filled]
\label{thm:maximal-volume}
Under the synchrony and epoch-duration assumptions of
\Cref{sec:protocol}, the extended protocol satisfies $f$-Maximal volume
filled as defined in \Cref{def:maximal-volume}.
\end{theorem}

\begin{proof}
Let $S$ be the set in the property definition. Every honest validator records
each $\tx\in S$ and includes it in its mini-block. Since $V^*$ contains at
least $M_B-f\geq n-2f>0$ honest validators, every request in $S$ appears in
$U$. Thus $S\subseteq U$.

Any admissible fill of $S$ extends to an admissible fill of $U$ by assigning
zero to $U\setminus S$. The added requests require no execution, and the
filled amount in the prefix of every positive-filled request is unchanged.
Hence its inequality in \Cref{def:partial-admissible} remains satisfied.
It follows that $R(U)\geq R(S)$.

By \Cref{lem:maximal-volume}, the protocol fills total volume $R(U)$.
Discarding zero-filled requests changes no volume, so its final block $T$
satisfies
\[
 \sum_{(\tx_i,\filled_i)\in T}\filled_i
 =R(U)\geq R(S).
\]
\qed
\end{proof}

\section{Combining Buys and Sells}

We extend our main protocol to allow for simultaneous buys and sells.
Each Validator receives trade requests with both buys and sells and must decide when it is able to pre-approve a request.
We add $\mathsf{side}\in\{\mathsf{buy},\mathsf{sell}\}$ to the
existing trade request format: $\tx=(\mathsf{side},\amt,\limit)$.
A naive way to handle both buys and sells would be to run two independent instances of the the protocol from \Cref{sec:protocol}, one for buys and one for sells. However, this approach approves would forego the opportunity of approving more buys when there are matching sells and conversely.

The challenge lies in determining when a validator can safely pre-approve additional buys based on the sell requests it has seen (and vice versa) without violating any limit prices.
Towards this, we first propose an extended definition of admissibility for the bilateral setting with both buys and sells (\Cref{def:bilateral-admissibility}).
Unlike the unilateral setting (\Cref{def:admissible}), bilateral admissibility involves two sets of trade requests: a set $G$ of pre-approved trade requests that must be executed fully and a set $U \supseteq G$ of trade requests that can optionally be executed to fill enough liquidity to support the trades in $G$.  
We show a pre-approval rule and block finalization protocol (similar to those in \Cref{sec:protocol}) that ensures bilateral admissibility when $G$ is the set of pre-approved trade requests and $U$ is the set of all other trade requests reported in any finalized block that may or may not execute.
Finally, we show in lemma \ref{lem:matching} that when the set of trades in a block is bilaterally admissible, all pre-approved trades in $G$ (and optionally some others in $U$) can be executed while satisfying all limit prices.

We extend our AMM curve $p(x)$ to allow for either positive or negative net purchases $x$  and write
$P(x)=\int_0^x p(u)\,du$. Let $c(\ell)$ be the inverse of $p(x)$, with $c(p(0))=0$ and both $c(.),p(.)$ continuous. Write $[z]_+=\max\{z,0\}$.
At limit $\ell$, the AMM offers supply $[c(\ell)]_+$ and demand
$[-c(\ell)]_+$.  

\begin{definition}[Bilateral admissibility]
\label{def:bilateral-admissibility}
Let $G \subseteq U$ be a collection of trade requests.
$G$ is said to be bilaterally admissible in $U$ if for every $\ell$,
\begin{align}
 \sum_{\substack{\tx_i\in G\text{ buys}\\\limit_i\leq\ell}}\amt_i
 &\leq [c(\ell)]_++
 \sum_{\substack{\tx_j\in U\text{ sells}\\\limit_j\leq\ell}}\amt_j,
 \label{eq:hallbuy}\\
 \sum_{\substack{\tx_i\in G\text{ sells}\\\limit_i\geq\ell}}\amt_i
 &\leq [-c(\ell)]_++
 \sum_{\substack{\tx_j\in U\text{ buys}\\\limit_j\geq\ell}}\amt_j.
 \label{eq:hallsell}
\end{align}
\end{definition}

In our extended protocol, 
similar to \Cref{sec:partial_fills_buy},
validators keep track of not only trade requests they pre-approved but also all trade requests that they have seen. An honest validator pre-approves only trade requests where it has sufficient capacity. As new trade requests are received that may increase capacity, it promotes old seen trade requests to 
pre-approved in the order received. When forming a mini-block, validators report separate lists of seen and pre-approved trade requests that can be used to build a final block.
Similar to \Cref{sec:partial_fills_buy}, trades included in the final block will have a parameter $\filled_i \in [0,\amt_i]$ denoting partial filling of trade requests.
In addition, each trade will also be assigned $\price_i$ denoting the execution price of the trade.
When we considered only buys, we had implcitly considered $\price_i$ to be the price determined by the AMM curve after trades has been ordered by their limit prices.
Our goal is to ensure that $\price_i \leq \limit_i$ for all executed buys and $\price_i \geq \limit_i$ for all executed sells.

\textbf{Modified Pre-Approval Protocol.}
Alongside the original pre-approval and certificate sets, validator $v$
records trade requests $\mathcal{R}_v$ that it has seen. Before the finalization cutoff, it records
and relays every valid trade request, including trade requests that fail the approval test.
  For each validator $v$, given the set of trade requests it has seen, define

\[
 \widehat S_v(\ell)=
 \sum_{\substack{\tx_i\in \mathcal{R}_v\text{ sells}\\\limit_i\leq\ell}}\amt_i,
 \qquad
 \widehat B_v(\ell)=
 \sum_{\substack{\tx_i\in \mathcal{R}_v\text{ buys}\\\limit_i\geq\ell}}\amt_i.
\]
Our validator's new rule for adding a trade request to its pre-approved set $A_v$ is that it will pre-approve if doing so does not violate
\begin{eqnarray}
\sum_{\substack{\tx_i\in\myapproved{v}\text{ buys}\\\limit_i\leq\ell}}\amt_i
 \leq\FR\,\bigl([c(\ell)]_++\widehat S_v(\ell)\bigr)
 \quad\text{for every }\ell \label{eq:buy} \\
\sum_{\substack{\tx_i\in\myapproved{v}\text{ sells}\\\limit_i\geq\ell}}\amt_i
 \leq\FR\,\bigl([-c(\ell)]_++\widehat B_v(\ell)\bigr)
 \quad\text{for every }\ell.
 \label{eq:sell}
\end{eqnarray}
Note that this is the analogue of $\FR$-admissibility (\Cref{def:fr-admissible}) for bilateral admissibility (\Cref{def:bilateral-admissibility}).

\textbf{Modified Finalization Protocol.}
In our modified finalization protocol, each mini-block $B_v = (\myapproved{v},\mycerts{v},\mathcal{R}_v)$ additionally contains the recorded requests $\mathcal{R}_v$. Well-formedness verifies these requests and replaces the original $\eta$-admissibility by checking \Cref{eq:buy,eq:sell}. The quorum requirement and approval-certificate cross-check for block validity are as before.
The inclusion and execution ordering steps are described below.
First, if $U$ is the set of all trade requests appearing in the block (including recorded and approved), note that all pre-approved transactions $G \subseteq U$ satisfy joint bilateral feasibility.

\Needspace{8\baselineskip}
\begin{lemma}[Admissibility of pre-approved requests]\label{thm:safety}
Under \eqref{eq:buy}--\eqref{eq:sell} with $\FR=(q-f)/q$, the set $G$ of
trade requests with at least $M_B-f$ approvals among selected mini-blocks is bilaterally admissible.
\end{lemma}
\begin{proof}
For each limit $\ell$, let $X(\ell)$ be the total quantity of buys in
$G$ with limits at most $\ell$, and let $X_v(\ell)$ be the
corresponding quantity in $\myapproved{v}$. Let $S_U(\ell)$ be the total
quantity of sells in $U$ with limits at most $\ell$.
Every trade request recorded as seen in a selected mini-block belongs to $U$, so
$\widehat S_v(\ell)\leq S_U(\ell)$ for every $v\in V^*$.
Counting approvals and summing the local buy budgets gives
\[
\begin{aligned}
 X(\ell)
 &\leq \frac{1}{M_B-f}\sum_{v\in V^*}X_v(\ell)
 \leq \frac{\FR M_B}{M_B-f}\bigl([c(\ell)]_++S_U(\ell)\bigr)
 \leq [c(\ell)]_++S_U(\ell).
\end{aligned}
\]
This demonstrates \Cref{eq:hallbuy} for buys and an analogous argument demonstrates \Cref{eq:hallsell} for sells.
\qed
\end{proof}

Next, we show how to compute the amounts filled for each trade request $\filled_i$, given a block with trades $U$ and pre-approved trades $G\subseteq U$ that are bilaterally admissible.
\Cref{lem:matching} shows that all trades in $G$ are completely filled, every filled trade can be assigned a price that satisfies its limit, and the AMM receives the correct revenue.
This is because the bilateral admissibility ensures that the trades in $U$ supply enough demand or supply to augment the AMM's own demand and supply to meet the supply or demand of the pre-approved trades in $G$.
It also preserves monotonicity: a trade with a higher limit price will not be executed at a lower price than a trade with a lower limit price (monotonicity was implicit in \Cref{sec:protocol} by ordering trades by limit prices).

Our protocol works in two steps.  First, we compute the quantity of buys minus the quantity of sells in the set of pre-approved trades $G$.  We then compute a price $\ell^*$ where executing all buys in $U\setminus G$ with limit prices below $\ell^*$ and all sells above $\ell^*$ (with potential partial fills at $\ell^*$) allows all trades in $G$ to execute.  This results in a total set $U^*$ of trades that actually execute.   If $U^*$ has more buys than sells so that the AMM is a net seller, sells in $U^*$ are effectively added to the AMM's liquidity supply.  Buys are executed in order of limit price, where each buy is greedily given the best possible price provided by the remaining sell liquidity as in our main protocol.   Sells execute at their limit price. If $U^*$ had more sells, we reverse this procedure so buys fill at their limit price and sells are given price improvement.  

This pricing preserves the property in our main protocol that a trade's execution price is an increasing function of its limit price.  Because requesting an attractive limit price lowers a trade request's chance of approval, this allows traders with a particularly high desire to trade to choose an unattractive limit price to increase their odds of acceptance. In our economic analysis, this allowed for a ``selection effect'' where traders who benefit the most from trading have the highest chance of their trades executing.

\begin{description}
\item[\textbf{Input:}] Set of seen trade requests $U$ and set of pre-approved transactions $G\subseteq U$, with $G$ bilaterally admissible in $U$.
\item[\textbf{Output:}] Fill vector $\filledvec=(\filled_i)_{\tx_i\in U}$
\item[Pre-approval inclusion.] Set $\filled_i=\amt_i$ for every request in $G$ and initialize all other fills to zero. Define the guaranteed net demand
\[
 g=\sum_{\substack{\tx_i\in G\text{ buys}}}\amt_i
 -\sum_{\substack{\tx_i\in G\text{ sells}}}\amt_i
\]
and the right-continuous augmented liquidity curve
\begin{equation}\label{eq:augmented-liquidity}
 \widetilde c(\ell)=c(\ell)
 +\sum_{\substack{\tx_i\in U\setminus G\text{ sells}\\\limit_i\leq\ell}}\amt_i
 -\sum_{\substack{\tx_i\in U\setminus G\text{ buys}\\\limit_i>\ell}}\amt_i.
\end{equation}
\item[Limit-matched-inclusion.] Choose the cutoff
$
 \ell^*=\inf\{\ell:\widetilde c(\ell)\geq g\}.
$
Among requests in $U\setminus G$, fully fill buys with $\limit_i>\ell^*$ and sells with $\limit_i<\ell^*$. Leave buys with $\limit_i<\ell^*$ and sells with $\limit_i>\ell^*$ unfilled.
\label{item:imbalance-fill} Define the remaining imbalance
\[
 \Delta_{\ell^*}=g-\widetilde c(\ell^*)
 +\sum_{\substack{\tx_i\in U\setminus G\text{ sells}\\\limit_i=\ell^*}}\amt_i.
\]
If $\Delta_{\ell^*}>0$, fill this quantity of sells with limit $\ell^*$; if $\Delta_{\ell^*}<0$, fill $-\Delta_{\ell^*}$ of buys with limit $\ell^*$. Process these requests in a fixed order by identifier, assigning each at most $\amt_i$ until the required quantity is reached. Leave all other requests at the cutoff unfilled.
\item Output $\filledvec=(\filled_i)_{\tx_i\in U}$.
\end{description}

\begin{lemma}[Limit-satisfying prices for bilateral admissibility]\label{lem:matching}
If $G\subseteq U$ is bilaterally admissible and $\filledvec$ is as computed above, then there exists $\pricevec$ such that:
\begin{enumerate}
  \item all pre-approved trades are completely filled: for all $\tx_i\in G$, $\filled_i = \amt_i$;
  \item all limits are respected: for all $\tx_i\in U$ such that $\filled_i > 0$, $\mathsf{side}_i = \mathsf{buy} \implies \price_i \leq \limit_i$ and $\mathsf{side}_i = \mathsf{sell} \implies \price_i \geq \limit_i$;
  \item the AMM's revenue is exactly $P(x)$ where $x := \sum_{\tx_i\in U \text{buys}}\filled_i - \sum_{\tx_i\in U \text{sells}}\filled_i$ is the net purchase quantity;
  \item prices are monotonic: for all $\tx_i, \tx_j \in U$ such that $\filled_i > 0$ and $\mathsf{side}_i = \mathsf{side}_j$, $\limit_i \leq \limit_j \implies \price_i \leq \price_j$.
\end{enumerate}
\end{lemma}
\begin{proof}
We first justify that the cutoff $\ell^*$ exists and is finite. For this argument, let
$a=\min(\{p(0)\}\cup\{\limit_i:\tx_i\in U\})$ and
$b=\max(\{p(0)\}\cup\{\limit_i:\tx_i\in U\})$.
Since $c(a)\leq0$ and every request has limit at least $a$, \Cref{eq:hallsell} at $a$ gives
\[
 \widetilde c(a^-)
 =c(a)-\sum_{\substack{\tx_i\in U\setminus G\text{ buys}}}\amt_i
 \leq g.
\]
Similarly, $c(b)\geq0$ and every request has limit at most $b$, so \Cref{eq:hallbuy} at $b$ gives
\[
 g\leq c(b)+\sum_{\substack{\tx_i\in U\setminus G\text{ sells}}}\amt_i
 =\widetilde c(b).
\]
For every $\ell<a$, strict monotonicity of $c=p^{-1}$ gives
\[
 \widetilde c(\ell)
 =c(\ell)-\sum_{\substack{\tx_i\in U\setminus G\text{ buys}}}\amt_i
 <\widetilde c(a^-)\leq g.
\]
Thus the set defining the cutoff is nonempty and has no element below $a$, so its infimum lies in $[a,b]$.
Second, we show that there are enough trades with limit exactly $\ell^*$ to fill the quantity $\pm\Delta_{\ell^*}$.
The curve $\widetilde c$ is nondecreasing and right-continuous, with only upward jumps. Consequently,
\[
 \widetilde c(\ell^*-)\leq g\leq\widetilde c(\ell^*).
\]
Here the left-hand value includes buys at the cutoff and excludes sells at the cutoff. Consequently,
\[
 -\sum_{\substack{\tx_i\in U\setminus G\text{ buys}\\\limit_i=\ell^*}}\amt_i
 \leq\Delta_{\ell^*}\leq
 \sum_{\substack{\tx_i\in U\setminus G\text{ sells}\\\limit_i=\ell^*}}\amt_i.
\]
There is therefore enough volume at the cutoff to fill $\pm\Delta_{\ell^*}$ while respecting every request's quantity bound.

\textbf{Price construction.}
Let $U^*$ be the set of executed requests. Their net asset purchase is
\[
 x=\sum_{\substack{\tx_i\in U^*\text{ buys}}}\filled_i
 -\sum_{\substack{\tx_i\in U^*\text{ sells}}}\filled_i
 =c(\ell^*).
\]
Define the AMM's executed volume with marginal price at most $\ell$ by
\[
F_{\mathrm{AMM}}(\ell)
=\left|\int_0^x \mathbf{1}_{\{p(v)\leq\ell\}}\,dv\right|.
\]
We now define two distribution functions that represent the set of buy and sell limit prices that that will execute as actual trades, including the liquidity supplied by the AMM.
\begin{align}
F_{U^*,\text{buy}}(\ell)
&=\sum_{\substack{\tx_i\in U^*\text{ buys}\\\limit_i\leq\ell}}\filled_i
+\mathbf{1}_{\{x<0\}}F_{\mathrm{AMM}}(\ell),\\
F_{U^*,\text{sell}}(\ell)
&=\sum_{\substack{\tx_i\in U^*\text{ sells}\\\limit_i\leq\ell}}\filled_i
+\mathbf{1}_{\{x>0\}}F_{\mathrm{AMM}}(\ell),
\end{align}
The total buy and sell quantities are equal, i.e., $F_{U^*,\text{buy}}(\infty) = F_{U^*,\text{sell}}(\infty) = M$, and all liquidity $F_{\mathrm{AMM}}(\ell)$ is only added on one side to meet the quantity imbalance in $U^*$ between buys and sells.

Define the inverse functions $F_{U^*,\text{buy}}^{-1}(u)$ with $F_{U^*,\text{sell}}^{-1}(u)$ where $F^{-1}(u)=\inf\{\ell:F(\ell)\ge u\}$.

If $x \geq 0$, assign $\price_i = \limit_i$ for each sell in $U^*$. Sort all buys in $U^*$ by limit prices, treating buys with a tied limit price as a single larger buy.
For the $i$-th buy, let $I_i := \left( \sum_{j=1}^{i-1} \filled_j, \sum_{j=1}^i \filled_j \right]$ be the interval it occupies in the sorted order.
Set the $i$-th buy's price as
\[
 \price_i=\frac{1}{\filled_i}\int_{I_i}F_{U^*,\text{sell}}^{-1}(u)\,du.
\]
If $x < 0$, assign $\price_i = \limit_i$ for each buy in $U^*$. Sort all sells in $U^*$ by limit prices, also merging tied limit prices into a single sell.
For the $i$-th sell, let $I_i := \left( \sum_{j=1}^{i-1} \filled_j, \sum_{j=1}^i \filled_j \right]$ be the interval it occupies in the sorted order.
Set the $i$-th sell's price as
\[
 \price_i=\frac{1}{\filled_i}\int_{I_i}F_{U^*,\text{buy}}^{-1}(u)\,du.
\]
A buyer pays $\filled_i\price_i$ and a seller receives $\filled_i\price_i$.

\textbf{Limit validity.}
If we can show that $F_{U^*,sell}^{-1}(u) \leq F_{U^*,\text{buy}}^{-1}(u)$, then executing at the above price satisfies limits.
This is because the $i$-th buy's price is exactly $\limit_i$ when $x < 0$ and when $x \geq 0$,
\[
  \price_i = \frac{1}{\filled_i}\int_{I_i}F_{U^*,\text{sell}}^{-1}(u)\,du \leq \frac{1}{\filled_i}\int_{I_i}F_{U^*,\text{buy}}^{-1}(u)\,du = \limit_i,
\]
Similarly, when $x < 0$, for the $i$-th sell,
\[
  \limit_i = \frac{1}{\filled_i}\int_{I_i}F_{U^*,\text{sell}}^{-1}(u)\,du \leq \frac{1}{\filled_i}\int_{I_i}F_{U^*,\text{buy}}^{-1}(u)\,du = \price_i.
\]

 We prove the equivalent statement that $F_{U^*,\text{buy}}(\ell)\leq F_{U^*,\text{sell}}(\ell)$ pointwise.  For $\ell<\ell^*$, executed buys with limits at most $\ell$ are exactly the corresponding buys in $G$, and all available sells with limits at most $\ell$ in $U$ fill completely. Since $c(\ell)\leq x=c(\ell^*)$, the definition of $F_{\mathrm{AMM}}$ gives
\[
\mathbf{1}_{\{x<0\}}F_{\mathrm{AMM}}(\ell)=0,
\qquad
\mathbf{1}_{\{x>0\}}F_{\mathrm{AMM}}(\ell)=[c(\ell)]_+.
\]
Thus inequality \eqref{eq:hallbuy} gives $F_{U^*,\text{buy}}(\ell)\leq F_{U^*,\text{sell}}(\ell)$. For $\ell\geq\ell^*$, we instead have $c(\ell)\geq x$, so
\[
\begin{aligned}
\mathbf{1}_{\{x>0\}}\bigl(|x|-F_{\mathrm{AMM}}(\ell)\bigr)&=0,\\
\mathbf{1}_{\{x<0\}}\bigl(|x|-F_{\mathrm{AMM}}(\ell)\bigr)&=[-c(\ell)]_+.
\end{aligned}
\]
Equal total mass and the cutoff rule for filling trades on each side of $\ell^*$ therefore give
\[
\begin{aligned}
F_{U^*,\text{buy}}(\ell)-F_{U^*,\text{sell}}(\ell)
&=
\sum_{\substack{\tx_j\in G\text{ sells}\\\limit_j>\ell}}\amt_j
-
\sum_{\substack{\tx_i\in U\text{ buys}\\\limit_i>\ell}}\amt_i\\
&\quad-\mathbf{1}_{\{x<0\}}\bigl(|x|-F_{\mathrm{AMM}}(\ell)\bigr)
\leq 0,
\end{aligned}
\]
where the inequality follows from \eqref{eq:hallsell} applied just above $\ell$. Thus $F_{U^*,\text{buy}}\leq F_{U^*,\text{sell}}$ everywhere.

\textbf{AMM payment and price monotonicity.}
If $x\geq0$, total buyer payments are
\[
 \int_0^M F_{U^*,\text{sell}}^{-1}(u)\,du
 =\sum_{\substack{\tx_i\in U^*\text{ sells}}}\filled_i\limit_i+P(x),
\]
since the sell distribution includes the AMM's supply at marginal prices $p(v)$ for $v\in[0,x]$. After paying trader sellers at their limits, the AMM retains $P(x)$. If $x<0$, total seller receipts are
\[
 \int_0^M F_{U^*,\text{buy}}^{-1}(u)\,du
 =\sum_{\substack{\tx_i\in U^*\text{ buys}}}\filled_i\limit_i-P(x),
\]
since the AMM's buy orders have total value $\int_x^0 p(v)\,dv=-P(x)$. Subtracting these receipts from trader buyers' payments again gives a net AMM receipt of $P(x)$.
Finally, both $F_{U^*,\text{buy}}^{-1}$ and $F_{U^*,\text{sell}}^{-1}$ are nondecreasing. Requests on one side execute at their limits; on the other side, execution prices are averages of the corresponding inverse function over consecutive intervals $I_i$. An earlier interval's average cannot exceed a later interval's average, so execution prices are nondecreasing on each side in the prescribed order of limits and identifiers.
\qed
\end{proof}

\textbf{Security properties.}
Price validity and pre-approval inclusion are implied by \Cref{lem:matching}.
Pre-approval liveness holds with $\FR$-admissibility replaced by \Cref{eq:buy,eq:sell}.
Agreement and accountable inclusion remain unchanged.

\textbf{LP deposits and withdrawals at block end.}
A standard extension to our protocol would allow passive LPs to deposit and withdraw liquidity in the AMM.  We impose that all deposits and withdrawals are handled at the end of a block, after regular trade requests have been executed.  Validators pre-approve all valid LP deposit and withdrawal requests, and such a pre-approval in even a single selected mini-block ensures execution of a valid request.  After all trades settle, process every distinct valid LP request. Let
$R=(R_A,R_C)$ be the post-trade reserves in the AMM and $\Lambda>0$ the outstanding
LP shares before processing these requests. All conversions use this
same snapshot:
\begin{itemize}
\item A withdrawal burns its locked $w$ shares and receives
$(w/\Lambda)R$.
\item A deposit with locked asset caps $(d_A,d_C)$ mints
\[
 m=\Lambda\min\{d_A/R_A,\ d_C/R_C\},
\]
contributes $(m/\Lambda)R$, and receives its unused escrow back. Any deposit that does not have sufficient inventory in both assets does not execute, and any withdrawal without enough shares does not execute.
\end{itemize}

\section{Related Work}
\label{sec:related}

\noindent\textbf{AMM arbitrage and loss-versus-rebalancing.}
Constant-function market makers (CFMMs) enable decentralized asset exchange through deterministic pricing curves~\cite{bancor2017,adams2018uniswap,angeris2020improved}. Because AMM prices change only when trades occur, movements in external market prices create arbitrage opportunities against stale AMM prices, resulting in \emph{loss-versus-rebalancing} (LVR)~\cite{milionis2022lvr}. Prior work studies the effects of fees and arbitrage on these losses and on liquidity provision~\cite{milionis2024fees,capponijia2025}, while recent empirical work highlights the role of large price movements in CEX--DEX arbitrage~\cite{ladoczki2026mev}. \projname keeps the underlying CFMM pricing rule but changes when traders receive an execution commitment. Traders can obtain a pre-approval while the block is still being formed rather than waiting until its end. Our economic analysis studies how this earlier commitment reduces the incentive to wait for a large CEX--DEX price discrepancy before trading.

\noindent \textbf{MEV mitigation and transaction ordering.}
A growing literature studies how execution and ordering rules can reduce maximal extractable value (MEV), including front-running, sandwich attacks, and arbitrage~\cite{daian2020flashboys,roughgarden2021transaction}. One approach is to reduce the importance of ordering. Batch execution~\cite{canidio2023batching} executes trades together at a uniform end-of-batch price, while SPEEDEX~\cite{ramseyer2023speedex} and Groundhog~\cite{groundhog} avoid imposing an intra-block order through commutative state updates; these systems still rely on consensus to determine end-of-batch outcomes. A complementary line seeks fairer sequencing: order fairness~\cite{kelkar2020orderfair} and  enforceable sequencing rules~\cite{gsr2024}. For AMMs specifically, CLVR~\cite{clvr2024} reduces intra-block price movements, while Chan, Wu, and Shi~\cite{chan2025mechanism} design an incentive-compatible mechanism requiring fair ordering. Diamandis and Angeris~\cite{diamandis2023welfare} show, however, that fair ordering need not yield economically efficient allocations.
Moreover, Condorcet cycles limit fair ordering.
Practical systems such as MEV Blocker~\cite{mevblocker2023} and SUAVE~\cite{suave2023} instead use order-flow auctions and alternative block-building mechanisms to reduce or redistribute ordering-based MEV. \projname 
deterministically executes pre-approved trades,
without the complex task of   fairly  ordering transactions by arrival time.

\noindent \textbf{Consensus-free fast paths and pre-confirmations.}
Several systems bypass full consensus for latency-sensitive operations. Prior work shows that payments and single-owner transactions can be confirmed by validator quorums without consensus~\cite{guerraoui19consensus,auvolat2020moneytransfer,kuznetsov2023assert-transfer,baudet2020fastpay,blackshear2024suilutris,linera2023}, while hybrid protocols optimistically avoid consensus and fall back to it upon conflicts~\cite{zzyzx,cowling2006hq}. Recent work shows that some shared-object operations can also avoid consensus. Stingray~\cite{stingray2024} uses local approvals against scaled capacity bounds, CryptoConcurrency~\cite{tonkikh2023cryptoconcurrency} avoids consensus when concurrent transfers are jointly feasible, and Zef~\cite{baudet2022defizef} explores consensus-free decentralized assets and swaps. \projname applies similar ideas to AMMs, where trades compete for  liquidity and independently approved trades may not all be  feasible. We use limit prices together with a scaled admissibility rule
to pre-approve trades while maintaining admissibility.
Recent work on pre-confirmations \cite{botha2026preconfirmations} makes commitments before block confirmation for censorship resistance. \projname obtains similar guarantees directly from a quorum of validators and provides a different economic rationale for their usefulness.

\section*{Acknowledgments}
SS was supported in part by the AFOSR Award FA9550-24-1-0156, and research grants from the Bakar Fund, Peder Sather Funds, the Stellar Development Foundation, Supra Inc., Category Labs, Input Output Global, and Byzantine Research Inc.

\bibliographystyle{splncs04}
\bibliography{references}

\appendix
\crefalias{section}{appendix}
\crefalias{subsection}{appendix}

\section{Brownian \projname{} equilibrium and outcome formulas}
\label{app:strategic-timing}

\input{econ-glossary}

This appendix proves the threshold formula~\eqref{eq:rationalbid} and derives the execution probability~\eqref{eq:auction-diffusion-fill}, welfare~\eqref{eq:auction-diffusion-welfare}, and AMM gains~\eqref{eq:auction-diffusion-gains} stated in \Cref{sec:auction}.

\paragraph{1. First-order condition.}
A trader with private value $\eps(r)$ chooses a cutoff $p_{\mathrm{ex}}^*(s)$,
at which a mass $s$ has traded and the AMM price is $p(s)$, to maximize
\begin{equation}
e^{-\gamma_{\mathrm{diff}}[p_{\mathrm{ex}}^*(s)-p_{\mathrm{ex},0}]}
[p_{\mathrm{ex}}^*(s)+\eps(r)-p(s)]. \label{eq:traderObj}
\end{equation}
Differentiating with respect to $s$ and evaluating at $s=r$ gives
\begin{equation}
\gamma_{\mathrm{diff}}[p_{\mathrm{ex}}^*(r)+\eps(r)-p(r)]
=1-\frac{p'(r)}{p_{\mathrm{ex}}^{*\prime}(r)}.
\end{equation}

\paragraph{2. Constant markup.}
For $p_{\mathrm{ex}}^*(r)=p(r)-\eps(r)+c$ and $p'(r)=-\kappa\eps'(r)$,
we have $p_{\mathrm{ex}}^{*\prime}(r)=-(1+\kappa)\eps'(r)$. The FOC becomes
\begin{equation}
\gamma_{\mathrm{diff}}c=1-\frac{\kappa}{1+\kappa}
\quad\Longrightarrow\quad
c=\frac{1}{\gamma_{\mathrm{diff}}(1+\kappa)},
\end{equation}
which gives~\eqref{eq:rationalbid}.  Because the derivative of the trader's objective function in \Cref{eq:traderObj} only crosses zero once, this choice of trading time is globally optimal.

Because total asset supply is fixed and passive liquidity is allocated independently of private values, the mechanism-dependent component of expected welfare is
\begin{equation}
\mathcal W_{\mathrm{FS}}^{\mathrm{diff}}=\int_0^1 \eps(r)\Pr\nolimits_{\mathrm{FS}}^{\mathrm{diff}}(r)\,dr.
\end{equation}
For AMM gains from trade, let $\tau_r$ be the first time $p_{\mathrm{ex},t}$ reaches $p_{\mathrm{ex}}^{*}(r)$. Conditional on $\tau_r<T_{\mathrm{blk}}$, exponential memorylessness and the strong Markov property imply
\begin{equation}
\E[p_{\mathrm{ex},T_{\mathrm{blk}}}\mid \tau_r<T_{\mathrm{blk}}]=p_{\mathrm{ex}}^{*}(r).
\end{equation}
Thus the expected AMM markout on a filled type-$r$ trade is $p(r)-p_{\mathrm{ex}}^{*}(r)$, giving
\begin{equation}
\mathcal R_{\mathrm{FS}}^{\mathrm{diff}}=\int_0^1 [p(r)-p_{\mathrm{ex}}^{*}(r)]\Pr\nolimits_{\mathrm{FS}}^{\mathrm{diff}}(r)\,dr.
\end{equation}
\section{Allocation representation after a jump}
\label{app:jump-allocation}

This appendix proves the allocation representation in \Cref{lem:jump-allocation-representation} of \Cref{sec:auction}.

\begin{proof}[Proof of \Cref{lem:jump-allocation-representation}]
Fix a mechanism $\mathsf m\in\{\mathrm{FS},\mathrm{FIFO},\mathrm{batch}\}$. Write $e$ for private value, $a(e):=\Pr\nolimits_{\mathsf m}^{\mathrm{jump}}(1-F_\eps(e))$ for its interim allocation probability, and $t(e)$ for expected payment. The expected utility of a trader given its type $e$ is
\[
u(e)=(p_{\mathrm{ex},\mathrm{jump}}+e)a(e)-t(e).
\]
Because each trader could "mimic" the choices of any other trader and obtain their expected payment and fill probabilities $a(e'),t(e')$, we have the first-order condition $u'(e)=a(e)$. So long as the distribution $F_\eps$ has support on types who do not want to trade,  $u(e)\to0$ as $e\to-\infty$, and hence
\[
t(e)-p_{\mathrm{ex},\mathrm{jump}}a(e)=ea(e)-\int_{-\infty}^e a(s)\,ds.
\]
Integrating over $F_\eps$ and applying Fubini gives
\[
\mathcal R_{\mathsf m}^{\mathrm{jump}}=\int\left[e-\frac{1-F_\eps(e)}{f_\eps(e)}\right]a(e)\,dF_\eps(e).
\]
Since $r=1-F_\eps(e)$ and $\eps'(r)=-1/f_\eps(\eps(r))$, this is the stated rank formula. The welfare formula follows directly because, conditional on $p_{\mathrm{ex},\mathrm{jump}}$, allocating the asset to type $e$ contributes the mechanism-dependent surplus $e$.
\qed
\end{proof}

\section{Single crossing and jump-volume comparisons}
\label{app:convex-amm-crossing}

This appendix proves the decompositions and the \projname{}--FIFO single-crossing conditions of
\Cref{lem:single-crossing-welfare}. The lemma statement appears in \Cref{sec:auction}.

\textbf{Proof of sufficient condition for single crossing in \Cref{lem:single-crossing-welfare}}
\begin{proof}
We compare the slopes of the two fill probabilities. If their ratio
crosses one only upward, the allocation difference first decreases and
then may increase, giving the desired single crossing.

For prices $z$, define
\[
w(z)=\frac1{p'(p^{-1}(z))},\qquad
\alpha(z)=-\frac{w'(z)}{w(z)}
=\frac{p''(p^{-1}(z))}{p'(p^{-1}(z))^2},\qquad
\mathcal H(z)=\frac1{\alpha(z)}.
\]
The AMM assumption gives
\begin{equation}
\mathcal H'(z)=2-\mathcal C(p^{-1}(z))\in[0,1],
\qquad \mathcal H\text{ is concave}.
\label{eq:convex-amm-reciprocal-curvature}
\end{equation}
On the common interior allocation range, write $\theta=p_{\mathrm{ex},\mathrm{jump}}+e$,
$p_{\mathrm{bid}}(e)=p_{\mathrm{bid},\mathrm{FS}}^{\mathrm{jump}}(r(e))$, and $\delta=\theta-p_{\mathrm{bid}}(e)>0$.
Here $p_{\mathrm{bid}}$ abbreviates $p_{\mathrm{bid}}(e)$, and $p_{\mathrm{bid}}'(e)$ denotes differentiation
with respect to private value, not rank. The queue formulas and the differentiated
first-price FOC give
\begin{equation}
\begin{aligned}
\frac{d}{de}\Pr\nolimits_{\mathrm{FIFO}}^{\mathrm{jump}}(r(e))&=\frac{w(\theta)}{\overline F_\eps(e)},&
\frac{d}{de}\Pr\nolimits_{\mathrm{FS}}^{\mathrm{jump}}(r(e))&=\frac{p_{\mathrm{bid}}'(e)w(p_{\mathrm{bid}})}{\overline F_\eps(e)},\\
p_{\mathrm{bid}}'(e)&=\frac{1+\delta h_\eps(e)}{2+\delta\alpha(p_{\mathrm{bid}})}.
\end{aligned}
\label{eq:convex-amm-allocation-derivatives}
\end{equation}
Thus, defining the allocation difference and its slope ratio by
\[
\begin{aligned}
\Delta(e)&=\Pr\nolimits_{\mathrm{FS}}^{\mathrm{jump}}(r(e))-\Pr\nolimits_{\mathrm{FIFO}}^{\mathrm{jump}}(r(e)),\\
\mathcal D(e)&=\frac{\frac{d}{de}\Pr\nolimits_{\mathrm{FS}}^{\mathrm{jump}}(r(e))}
{\frac{d}{de}\Pr\nolimits_{\mathrm{FIFO}}^{\mathrm{jump}}(r(e))}
=\frac{w(p_{\mathrm{bid}})}{w(\theta)}\frac{1+\delta h_\eps(e)}{2+\delta\alpha(p_{\mathrm{bid}})}.
\end{aligned}
\]
we have $\Delta'(e)=w(\theta)[\mathcal D(e)-1]/\overline F_\eps(e)$.
All derivatives of $\Delta$ and $\mathcal D$ below are with respect to $e$.

We show that $\mathcal D'>0$ whenever $\mathcal D=1$. At such a point, set
\[
a=\delta\alpha(p_{\mathrm{bid}})>0,\qquad
k=p_{\mathrm{bid}}'(e)=\frac{w(\theta)}{w(p_{\mathrm{bid}})},\qquad
N=1+\delta h_\eps(e)=k(2+a),\qquad \beta=\mathcal H'(p_{\mathrm{bid}})\in[0,1].
\]
Since $N\ge1$ and $\mathcal H'\le1$,
\begin{equation}
\frac1{2+a}\le k
=\exp\left\{-\int_{p_{\mathrm{bid}}}^\theta\alpha(z)\,dz\right\}
\le\frac1{1+a}.
\label{eq:convex-amm-critical-ratio-bounds}
\end{equation}
Differentiating $\mathcal D$ at this point yields
\begin{align}
\delta N\frac{\mathcal D'}{\mathcal D}
={}&(1-k)(2k-1)+\delta^2h_\eps'(e)-\delta^2\alpha'(p_{\mathrm{bid}})k^2
+\delta N[\alpha(\theta)-\alpha(p_{\mathrm{bid}})k].
\label{eq:convex-amm-critical-derivative}
\end{align}
Here derivatives of $\alpha$ are with respect to price. Increasing hazard
gives $h_\eps'\ge0$, while concavity of $\mathcal H$ gives
$\alpha(\theta)\ge\alpha(p_{\mathrm{bid}})/(1+\beta a)$ and
$-\delta^2\alpha'(p_{\mathrm{bid}})=\beta a^2$. Therefore
\[
\delta N\frac{\mathcal D'}{\mathcal D}
\ge (1-k)(2k-1)+\beta a^2k^2
+ak(2+a)\left[\frac1{1+\beta a}-k\right].
\]
The right side decreases in $\beta$: its derivative is
$a^2k[k-(2+a)/(1+\beta a)^2]<0$ by
\eqref{eq:convex-amm-critical-ratio-bounds}. Setting $\beta=1$ hence gives
\[
\delta N\frac{\mathcal D'}{\mathcal D}\ge
L_a(k):=-1+\left[3+\frac{a(2+a)}{1+a}\right]k-2(1+a)k^2.
\]
This quadratic is concave, and at the endpoints of the permitted interval,
\[
L_a\!\left(\frac1{2+a}\right)
=\frac{a}{(2+a)^2(1+a)}>0,\qquad
L_a\!\left(\frac1{1+a}\right)=\frac{a}{(1+a)^2}>0.
\]
Thus $\mathcal D'>0$ whenever $\mathcal D=1$, so the slope ratio can cross one only upward.

\qed
\end{proof}

\section{Small- and large-jump comparisons}
\label{app:batch-sting-jump}

This appendix proves \Cref{lem:jump-volume-welfare-rankings}(ii) and (iii) of \Cref{sec:auction}.
We first prove part (ii), without linking the AMM curve to the private-value quantiles.
At $\Delta p_{\mathrm{mid}}=0$, batch fills exactly the positive-private-value traders.
For small positive $\Delta p_{\mathrm{mid}}$, its welfare loss is quadratic, while \projname{}
continues to execute a positive fraction of central negative types.

\begin{theorem}[Batch dominance near efficient clearing]
\label{thm:batch-sting-jump}
Suppose $f_\eps>0$ is $C^1$ on $\mathbb R$, symmetric about zero, and
log-concave, and $p\in C^2$ satisfies $p'>0$ and $p''\ge0$ on an
inventory domain containing $[0,1/2]$, with $p(0)$ finite.
Fix any equilibrium of the first-price jump game, and assume both
mechanisms have finite terminal prices and inventories inside this
domain. Set $\Delta p_{\mathrm{mid}}=p_{\mathrm{ex},\mathrm{jump}}-p(1/2)$, $s_p=p'(1/2)$,
$\Delta_p=p(1/2)-p(1/4)$, and $s_\eps=1/(2f_\eps(0))$.
If $\Delta_p\ge2F_\eps^{-1}(3/4)$ and
$0\le \Delta p_{\mathrm{mid}}<(s_\eps+s_p/4)\sqrt{\Delta_p/(24s_p)}$, then batch has strictly
higher welfare and AMM gains: $\mathcal W_{\mathrm{batch}}^{\mathrm{jump}}>\mathcal W_{\mathrm{FS}}^{\mathrm{jump}}$ and $\mathcal R_{\mathrm{batch}}^{\mathrm{jump}}>\mathcal R_{\mathrm{FS}}^{\mathrm{jump}}$.
\end{theorem}

\begin{proof}
Put $a=F_\eps^{-1}(3/4)$ and let $\mathcal W^*=\int_0^\infty e f_\eps(e)\,de$,
the welfare from filling exactly the positive-private-value traders.
Symmetric log-concavity implies $f_\eps(e)\le f_\eps(0)$ and a nondecreasing
hazard rate $h_\eps=f_\eps/(1-F_\eps)$. In particular,
\[
a\ge\frac1{4f_\eps(0)}=\frac{s_\eps}{2},\qquad
f_\eps(a)=\tfrac14h_\eps(a)\ge\tfrac14h_\eps(0)=\frac{f_\eps(0)}2.
\]
Since $\Delta_p\ge2a\ge s_\eps$ and $\Delta_p\le s_p/4$, we have
$s_\eps\le s_p/4$. The assumed upper bound on $\Delta p_{\mathrm{mid}}$ is therefore at most
$(s_p/2)\sqrt{1/96}<s_p/4$. We will prove the bound
\begin{equation}
\mathcal W_{\mathrm{batch}}^{\mathrm{jump}}-\mathcal W_{\mathrm{FS}}^{\mathrm{jump}}
\ge s_\eps\left[\frac{\Delta_p}{96s_p}
-\frac14\left(\frac{\Delta p_{\mathrm{mid}}}{s_\eps+s_p/4}\right)^2\right]>0.
\label{eq:batch-sting-gap}
\end{equation}

\textit{Batch's loss is quadratic.}
Write $-z$ for batch's marginal private value. Since $\Delta p_{\mathrm{mid}}\ge0$, we have
$z\ge0$, and symmetry and batch clearing give
\[
x_{\mathrm{batch}}^{\mathrm{jump}}=F_\eps(z),\qquad
\Delta p_{\mathrm{mid}}=z+p(F_\eps(z))-p(1/2)
\ge z+s_p[F_\eps(z)-1/2].
\]
For $\Delta p_{\mathrm{mid}}\le s_p/4$, this implies $F_\eps(z)\le3/4$ and hence $z\le a$.
The density is nonincreasing on $[0,a]$, so
$F_\eps(z)-1/2\ge z f_\eps(a)\ge z f_\eps(0)/2$. Therefore
\[
z\le\frac{\Delta p_{\mathrm{mid}}}{1+s_p f_\eps(0)/2},
\]
and
\begin{equation}
\mathcal W^*-\mathcal W_{\mathrm{batch}}^{\mathrm{jump}}
=\int_0^z e f_\eps(e)\,de
\le\frac{f_\eps(0)z^2}{2}
\le\frac{s_\eps}{4}\left(\frac{\Delta p_{\mathrm{mid}}}{s_\eps+s_p/4}\right)^2.
\label{eq:batch-efficient-loss}
\end{equation}

\paragraph{\projname{} has a uniform selection loss.}
A bid of $p(1/4)$ fills with probability at least $1/4$: the eligible
population has mass at most one, so cumulative
sales satisfy $x'(t)\le1$ and $x(t)\le t$, and an order arriving before
queue time $1/4$ finds the quote at or below $p(1/4)$.
For every private value $e\in[-a,0]$ this deviation gives
\begin{equation}
U_{\mathrm{FS}}^{\mathrm{jump}}(e)\ge\frac{\Delta p_{\mathrm{mid}}+e+\Delta_p}{4}
\ge\frac{\Delta_p-a}{4}\ge\frac{\Delta_p}{8}>0.
\label{eq:batch-sting-deviation}
\end{equation}
Let $p_{\mathrm{bid}}(e)$ denote the equilibrium bid of type $e$ and
$\delta(e):=p_{\mathrm{ex},\mathrm{jump}}+e-p_{\mathrm{bid}}(e)$ its surplus per fill, so that
$U_{\mathrm{FS}}^{\mathrm{jump}}(e)=\Pr\nolimits_{\mathrm{FS}}^{\mathrm{jump}}(1-F_\eps(e))\,\delta(e)$.
Since $U_{\mathrm{FS}}^{\mathrm{jump}}(e)>0$, the equilibrium bid fills with positive
probability, which requires $p_{\mathrm{bid}}(e)\ge p(0)$ because the AMM quote
never falls below $p(0)$. Hence
\[
\delta(e)\le p_{\mathrm{ex},\mathrm{jump}}+e-p(0)
\le \Delta p_{\mathrm{mid}}+\big[p(1/2)-p(0)\big]
\le\frac{s_p}{4}+\frac{s_p}{2}=\frac{3s_p}{4},
\]
using $e\le0$, the maintained restriction $\Delta p_{\mathrm{mid}}\le s_p/4$, and convexity
($p'\le s_p$ on $[0,1/2]$). Therefore
\[
\Pr\nolimits_{\mathrm{FS}}^{\mathrm{jump}}(1-F_\eps(e))=\frac{U_{\mathrm{FS}}^{\mathrm{jump}}(e)}{\delta(e)}
\ge\frac{\Delta_p/8}{3s_p/4}=\frac{\Delta_p}{6s_p},
\qquad -a\le e\le0.
\]
The argument uses only that type $e$'s equilibrium payoff weakly
exceeds the payoff from the deviation; no monotonicity or
differentiability of the equilibrium is needed.

The central negative types have a uniformly positive private-value
loss. By symmetry and $f_\eps\le f_\eps(0)$,
\[
\int_{-a}^0(-e)f_\eps(e)\,de
=\int_0^a\left[\frac14-(F_\eps(u)-1/2)\right]du
\ge\int_0^{s_\eps/2}\left[\frac14-f_\eps(0)u\right]du
=\frac{s_\eps}{16}.
\]
Consequently,
\[
\mathcal W^*-\mathcal W_{\mathrm{FS}}^{\mathrm{jump}}
=\int_0^\infty e[1-\Pr\nolimits_{\mathrm{FS}}^{\mathrm{jump}}(1-F_\eps(e))]f_\eps(e)\,de
+\int_{-\infty}^0(-e)\Pr\nolimits_{\mathrm{FS}}^{\mathrm{jump}}(1-F_\eps(e))f_\eps(e)\,de
\ge\frac{\Delta_p}{6s_p}\cdot\frac{s_\eps}{16}
=\frac{s_\eps\,\Delta_p}{96\,s_p}.
\]
Subtracting \eqref{eq:batch-efficient-loss} proves
\eqref{eq:batch-sting-gap}.

\textit{AMM gains are also higher.}
Let $\mathcal U_{\mathsf m}^{\mathrm{jump}}=\mathcal W_{\mathsf m}^{\mathrm{jump}}-\mathcal R_{\mathsf m}^{\mathrm{jump}}$ denote aggregate trader utility.
If $x_{\mathrm{FS}}^{\mathrm{jump}}\le x_{\mathrm{batch}}^{\mathrm{jump}}$, the batch price
$p_{\mathrm{batch}}^{\mathrm{jump}}=p(x_{\mathrm{batch}}^{\mathrm{jump}})$ is at least \projname{}'s terminal quote.
An infinitesimal trader can therefore obtain its batch payoff under
\projname{} by bidding $p_{\mathrm{batch}}^{\mathrm{jump}}$, which guarantees execution, if its value
is at least $p_{\mathrm{batch}}^{\mathrm{jump}}$, and by abstaining otherwise.
Optimality implies $\mathcal U_{\mathrm{FS}}^{\mathrm{jump}}\ge\mathcal U_{\mathrm{batch}}^{\mathrm{jump}}$.

If $x_{\mathrm{FS}}^{\mathrm{jump}}\ge x_{\mathrm{batch}}^{\mathrm{jump}}$, \Cref{lem:jump-allocation-representation}
gives $\mathcal U_{\mathsf m}^{\mathrm{jump}}=\int_0^1 \omega(r)\Pr\nolimits_{\mathsf m}^{\mathrm{jump}}(r)\,dr$, where
$\omega(r)=-r\eps'(r)=1/h_\eps(\eps(r))$ is positive and nondecreasing
because the hazard rate $h_\eps$ is nondecreasing.
At a fixed volume this integral is minimized by allocating to
the smallest ranks. Hence
\[
\mathcal U_{\mathrm{FS}}^{\mathrm{jump}}
\ge\int_0^{x_{\mathrm{FS}}^{\mathrm{jump}}}\omega(r)\,dr
\ge\int_0^{x_{\mathrm{batch}}^{\mathrm{jump}}}\omega(r)\,dr
=\mathcal U_{\mathrm{batch}}^{\mathrm{jump}}.
\]
In either case,
\[
\mathcal R_{\mathrm{batch}}^{\mathrm{jump}}-\mathcal R_{\mathrm{FS}}^{\mathrm{jump}}
=\mathcal W_{\mathrm{batch}}^{\mathrm{jump}}-\mathcal W_{\mathrm{FS}}^{\mathrm{jump}}+\mathcal U_{\mathrm{FS}}^{\mathrm{jump}}-\mathcal U_{\mathrm{batch}}^{\mathrm{jump}}
\ge \mathcal W_{\mathrm{batch}}^{\mathrm{jump}}-\mathcal W_{\mathrm{FS}}^{\mathrm{jump}}>0.
\]
\qed
\end{proof}

The condition $\Delta_p\ge2a$ says that the AMM's price rise from
inventory $1/4$ to $1/2$ exceeds the interquartile range of private
values. It guarantees that central negative types can profitably bid
for early liquidity. The jump bound keeps batch's marginal private
value close to zero, limiting its welfare loss to a quadratic in $\Delta p_{\mathrm{mid}}$.
The result applies to general convex AMMs, including constant-product
curves with sufficient liquidity, and uses no bound on the AMM slope
near inventory exhaustion. It also places no restriction on the \projname{}
equilibrium beyond existence: the welfare proof uses only that a
filled order pays at least $p(0)$ and that no type can gain by
switching to the bid $p(1/4)$; the AMM-gain comparison uses optimal
bidding and the envelope identity of
\Cref{lem:jump-allocation-representation}.

\begin{proof}[Proof of \Cref{lem:jump-volume-welfare-rankings}(iii)]
Write $\overline F_\eps=1-F_\eps$ and let $c_{\mathrm{FS}}^{\mathrm{jump}}$ be \projname{}'s terminal
eligibility cutoff. The limiting queue FOC gives
\[
c_{\mathrm{FS}}^{\mathrm{jump}}=p(x_{\mathrm{FS}}^{\mathrm{jump}})-p_{\mathrm{ex},\mathrm{jump}}
+p'(x_{\mathrm{FS}}^{\mathrm{jump}})\overline F_\eps(c_{\mathrm{FS}}^{\mathrm{jump}}).
\]
If $c_{\mathrm{FS}}^{\mathrm{jump}}>0$, symmetry gives $\overline F_\eps(c_{\mathrm{FS}}^{\mathrm{jump}})<1/2$.
Since $x_{\mathrm{FS}}^{\mathrm{jump}}\le1$, convexity and the jump condition imply
\[
c_{\mathrm{FS}}^{\mathrm{jump}}\le p(1)-p_{\mathrm{ex},\mathrm{jump}}+\frac12p'(1)\le0,
\]
a contradiction. Thus \projname{} fills every nonnegative-private-value
trader with certainty. FIFO does too, because its terminal quote is
$p(x_{\mathrm{FIFO}}^{\mathrm{jump}})\le p(1)<p_{\mathrm{ex},\mathrm{jump}}$. The single crossing can therefore be
chosen with $e^\star\le0$.

\projname{} bids do not exceed values, so its eligible mass at any common
inventory is at most FIFO's, giving $x_{\mathrm{FS}}^{\mathrm{jump}}\le x_{\mathrm{FIFO}}^{\mathrm{jump}}$ by queue
comparison. Both terms in the welfare decomposition of
\Cref{lem:single-crossing-welfare} are consequently nonnegative.
For AMM gains, \Cref{lem:jump-allocation-representation} replaces
private value $e$ by the virtual value $e-\overline F_\eps(e)/f_\eps(e)$.
The increasing hazard rate makes this function strictly increasing,
and its value at $e^\star\le0$ is negative. The same centered
selection--volume decomposition therefore gives nonnegative terms
for AMM gains. Both rankings are strict if allocations differ on a
set of positive measure.
\qed
\end{proof}

\section{Validator-count equivalence after a simultaneous jump}
\label{app:validator-count}

\subsection{Fair ordering: anonymity, neutrality, and a uniform queue}
\label{app:fair-validator-count}

Fix a simultaneous cohort of $m$ labeled orders. Conditional on all
submitted bids, quantities, and private values, suppose the $n$ validators'
receipt rankings are independent uniform permutations:
\[
  \pi_1,\ldots,\pi_n
  \stackrel{\mathrm{iid}}{\sim}\mathrm{Unif}(\mathfrak S_m).
\]
Independent continuous, identically distributed receipt delays satisfy
this assumption. We consider an aggregation rule $\mathcal A_n$ that
returns a complete ordering and satisfies the following axioms.
\begin{enumerate}
\item \emph{Transaction neutrality.} For every relabeling
$\rho\in\mathfrak S_m$,
\[
 \mathcal A_n(\rho\pi_1,\ldots,\rho\pi_n)
 \ \stackrel{d}{=}\
 \rho\,\mathcal A_n(\pi_1,\ldots,\pi_n).
\]
Here $\rho\pi$ relabels the entries of the ordered list $\pi$.
\item \emph{Validator anonymity.} Permuting the $n$ input rankings
does not change the output distribution.
\item \emph{Bid blindness and neutral resolution.} The aggregator uses
only the receipt rankings and fresh independent randomness. Bids,
values, quantities, and privileged transaction labels cannot affect
priority; any ties or partial orders are resolved into a complete order
in a transaction-neutral way.
\end{enumerate}

\begin{theorem}[Ideal fair ordering is equivalent to one random queue]
\label{thm:fair-validator-count}
Under these assumptions, for every $n\ge1$ and every conditional bid and
value profile,
\[
  \Pr\{\mathcal A_n(\pi_1,\ldots,\pi_n)=\sigma\}
  =\frac{1}{m!},
  \qquad \sigma\in\mathfrak S_m.
\]
Consequently, applying the same serial AMM execution and payment rule
to the aggregated queue gives exactly the same joint allocation and
payment law as one validator with a uniform receipt queue.
\end{theorem}

\begin{proof}
Condition on a fixed bid and value profile. The joint receipt-ranking
law is unchanged by a common relabeling $\rho$. Transaction neutrality
therefore implies that the output law is also unchanged by $\rho$.
For any two full rankings $\sigma$ and $\sigma'$, there is a relabeling
$\rho$ with $\rho\sigma=\sigma'$. Thus every ranking has the same
probability. Their probabilities sum to one, proving the displayed
formula. Bid blindness ensures that this argument remains valid after
conditioning on the traders' chosen bids.
\qed
\end{proof}

The stated axioms are satisfied, for example, by the Kemeny \citep{kemeny1959mathematics} and Ranked Pairs methods \citep{tideman1987independence,ramseyer2026fair} of aggregating preference orderings.

\subsection{\projname{} with bid-ordered AMM execution}
\label{app:fs-validator-count}

We now specify conditions under which any monotone equilibrium of the FS protocol yields the same allocation as in the version with a single centralized validator.   Let
\[
 \theta(r)=p_{\mathrm{ex},\mathrm{jump}}+\eps(r),
 \qquad
 D:=1-F_\eps\!\left(p(0)-p_{\mathrm{ex},\mathrm{jump}}\right)\in(0,1).
\]
 
\begin{proof}
\textit{Entry, execution prices, and the incentive equation.}
Every execution price is at least $p(0)$, so values below $p(0)$
cannot profit from entry. Conversely, any value above $p(0)$ can bid
strictly between $p(0)$ and its value. Arriving sufficiently early at
all validators and receiving a sufficiently early tie-break among
the highest signature scores has positive probability and gives
a feasible fill at a price no larger than this bid. Thus a regular
equilibrium cannot exclude an interval of values above $p(0)$.
Participation is precisely $r<D$, up to the indifferent boundary type.

Because bids strictly decrease with $r$, lower-value active types
execute first. Hence
\[
 X_n(r)=\int_r^D a_n(t)\,dt,\qquad
 p_n(r)=p(X_n(r)).
\]
A trader of true rank $r$ who submits rank $t$'s bid receives utility
\[
 a_n(t)\,[\theta(r)-p(X_n(t))].
\]
The first-order condition at $t=r$, together with $X_n'=-a_n$,
gives the following system, suppressing the validator subscript:
\begin{equation}
 \begin{split}
 X'(r)&=-a(r),\\
 a'(r)&=-\frac{a(r)^2p'(X(r))}{\theta(r)-p(X(r))},
 \qquad
 a(0+)=1,\quad X(D-)=0.
 \end{split}
 \label{eq:fs-validator-allocation}
\end{equation}
The boundary $X(D-)=0$ follows from the definition of execution depth.
A bid covering $p(2D)$
receives every signature and can be added to any feasible allocation,
so it guarantees execution at a price at most $p(D)$. Comparing this
deviation with an equilibrium payoff at most
$a_n(r)[\theta(r)-p(0)]$, and using $\theta(r)\to\infty$ as
$r\downarrow0$, proves $a_n(0+)=1$.  We show that there is at most one solution to this system of differential equations with these boundary conditions.

To prove uniqueness, set $Z=-a$. The system for $(X,Z)$ is a so-called cooperative system with $\frac{\partial X'}{ \partial Z}$ and $\frac{\partial Z'}{ \partial X}$ both positive:
\[
 X'=Z,\qquad Z'=\frac{Z^2p'(X)}{\theta-p(X)},\qquad
 \frac{\partial}{\partial X}
 \frac{p'(X)}{\theta-p(X)}
 =
 \frac{p''(X)[\theta-p(X)]+p'(X)^2}
      {[\theta-p(X)]^2}>0.
\]
By the comparison theorem for cooperative systems, a larger initial condition $X(0+)$ and the same $Z(0)=-1$ give
larger $X$ and weakly larger $Z$ pointwise. The difference between the two
$X$ paths is nondecreasing, so stays at least their strictly positive
initial difference. They cannot both end at $X(D-)=0$.
There is therefore at most one solution satisfying $a(0+)=1$ and $X(D-)=0.$

Thus every regular separating equilibrium, whenever it exists, has
\begin{equation}
 a_n(r)=a(r),\qquad
 X_n(r)=X(r),\qquad
 p_n(r)=p(X(r)),\quad 0<r<D,
 \label{eq:fs-validator-equivalence}
\end{equation}
where $p_n(r)$ is the expected price conditional on filling. 
\qed
\end{proof}

\section{Supplemental Protocol Information: Glossary, Algorithms and Proofs}
\label{app:algorithms}

\input{glossary}
 
This appendix gives pseudocode for the decentralized protocol of
\Cref{sec:protocol} and overflow security proofs.

\Cref{alg:trader-reservation} is run by
traders, \Cref{alg:validator-reservation} by validators during the
pre-approval stage, and \Cref{alg:block-consensus} by validators at
the end of the epoch. All parameters, the pricing function, and the notions of admissibility and $\FR$-admissibility are as defined in the main text. The
pseudocode describes a single epoch $e$. All local state is reset at the start
of each epoch, and approvals are signed together with the epoch number, so an
approval cannot be reused in a later epoch.
 
\begin{algorithm}[t]
\caption{Trader-side pre-approval protocol for trade request $\tx$ in epoch $e$}
\label{alg:trader-reservation}
\begin{algorithmic}[1]
\REQUIRE Trade request $\tx=(\amt,\limit)$
\STATE Send $\tx$ to all validators
\STATE $\mathsf{Responses} \gets \emptyset$
\WHILE{the pre-approval stage of epoch $e$ has not ended}
    \STATE Wait for a signed validator response, retransmitting $\tx$ to
    validators that have not responded
    \IF{the response is a valid signature on $\langle \mathsf{APPROVE}, \tx, e\rangle$
        from a validator not yet in $\mathsf{Responses}$}
        \STATE Add the response to $\mathsf{Responses}$
        \IF{$|\mathsf{Responses}| = q$}
            \STATE Form $\AC{\tx}$ from $\mathsf{Responses}$ and send it to all validators
            \RETURN $\AC{\tx}$
        \ENDIF
    \ENDIF
\ENDWHILE
\RETURN $\bot$
\end{algorithmic}
\end{algorithm}

\begin{algorithm}[t]
\caption{Validator-side processing at validator $v$ in epoch $e$}
\label{alg:validator-reservation}
\begin{algorithmic}[1]
\item[] \textbf{Upon receiving} trade request $\tx=(\amt,\limit)$:
\IF{$v$ has previously responded to $\tx$}
    \RETURN previously signed response
\ENDIF
\IF{the pre-approval stage of epoch $e$ has ended}
    \STATE Ignore $\tx$
\ENDIF
 
\item[]\vspace{0.3\baselineskip}
\COMMENT{Check $\FR${-}admissibility of $\myapproved{v} \cup \{\tx\}$; only limits $\ge \limit$ are affected}
\FORALL{$\ell \in \{\limit\} \cup \{\limit_i : \tx_i \in \myapproved{v},\ \limit_i > \limit\}$}
    \IF{$\ell < p((\sum_{\tx_j \in \myapproved{v}:\,\limit_j \le \ell} \amt_j/\FR)+ \amt/\FR)$}
        \RETURN signed $\langle \mathsf{REJECT},\tx,e\rangle$ \COMMENT{stored and replayed}
    \ENDIF
\ENDFOR
\STATE $\sigma_v(\tx) \gets$ signed $\langle \mathsf{APPROVE},\tx,e\rangle$
\STATE $\myapproved{v} \gets \myapproved{v} \cup \{(\tx, \sigma_v(\tx))\}$
\RETURN $\sigma_v(\tx)$
 
\item[]\vspace{0.3\baselineskip}
\item[] \textbf{Upon receiving} $\AC{\tx}$ before the end of the pre-approval stage:
\IF{$\AC{\tx}$ holds valid epoch-$e$ approvals of $\tx$ from $q$ distinct validators}
    \STATE $\Pi_v \gets \Pi_v \cup \{\AC{\tx}\}$
\ENDIF
\end{algorithmic}
\end{algorithm}

\begin{algorithm}[t]
\caption{End-of-epoch block consensus at validator $v$}
\label{alg:block-consensus}
\begin{algorithmic}[1]
\STATE \textbf{External validity predicate} (\Cref{def:valid}).
$\mathsf{Valid}(\mathcal B)=1$ if and only if:
\STATE \hspace{\algorithmicindent}
(i) $\mathcal B$ contains well-formed mini-blocks from at least $n{-}f$
distinct validators, where $B_u=(\myapproved{u},\Pi_u)$ is well-formed if
$\myapproved{u}$ is $\FR${-}admissible with every approval correctly signed
by $u$ for epoch $e$, and every $\AC{\tx}\in\Pi_u$ is correctly formed; and
\STATE \hspace{\algorithmicindent}
(ii) for every $\AC{\tx}$ in $\mathcal B$ and every $B_u$ in $\mathcal B$,
if $u$'s signature appears in $\AC{\tx}$, then the same signed approval of
$\tx$ appears in $\myapproved{u}$.
 
\vspace{0.5em}
\STATE \textbf{Protocol.}
\STATE At the end of the epoch, form the mini-block
$B_v \gets (\myapproved{v},\Pi_v)$ and sign it
\STATE Invoke MVBA with input $B_v$ and external-validity predicate $\mathsf{Valid}$
\STATE Let $\mathcal B$ be the decision by MVBA, $V^*$ the creators of its
mini-blocks, and $M_B \gets |V^*|$
\STATE \textbf{Pre-approved inclusion:} include every $\tx$ with approval
signatures from at least $\FR M_B$ distinct validators in $V^*$
\STATE \textbf{AC inclusion:} include every $\tx$ for which a valid
$\AC{\tx}$ appears in $\mathcal B$, unless already included
\STATE \textbf{Execution ordering:} order all included trade requests in
non-decreasing order of $\limit$, breaking ties deterministically by trade
request identifier
\STATE Output the resulting ordered block for execution
\COMMENT{optionally supplemented with non-pre-approved trades as in \Cref{sec:partial_fills_buy}}
\end{algorithmic}
\end{algorithm}

\subsection*{Proof of Agreement}

\begin{theorem}[$f$-Agreement]
\label{thm:agreement}
The protocol satisfies $f$-Agreement (defined in~\Cref{sec:security}).
\end{theorem}

\begin{proof}
By MVBA agreement, all honest validators decide the same output
$\mathcal B$. They apply the same inclusion and execution rules, including
the deterministic tie-breaking rule, and therefore output the same final
block.
\qed
\end{proof}

\subsection*{Proof of Accountable Inclusion.}

\begin{theorem}[Accountable inclusion]
\label{thm:accountable-inclusion}
The protocol satisfies accountable inclusion.
\end{theorem}

\begin{proof}
Let the evidence $w$ consist of an AC for $\tx$, the decided output
$\mathcal B$, $V^*$ the creators of its mini-blocks, and $M_B \gets |V^*|$. 
Define $J(w)$ to contain every
validator $v$ such that $v\in V^*$, the AC contains $v$'s approval of $\tx$,
and $\myapproved{v}$ omits that signed approval.

Suppose $\tx$ obtains an AC but is absent from the final block. The AC cannot
appear in any selected $\mycerts{v}$, because that would include $\tx$ directly.
Moreover, fewer than $\FR M_B$ selected approved sets contain $\tx$, because
otherwise the pre-approved inclusion rule would include it. The AC signer set
and $V^*$ intersect in at least $q+M_B-n\geq\FR M_B$ validators, by the bound
in the proof of \Cref{lem:representation}. Thus at least one signer in this
intersection omitted its approval, so $J(w)$ is nonempty. An honest validator retains every approval it issues in $\myapproved{v}$.
Therefore no valid evidence causes $J$ to identify an honest validator.
\qed
\end{proof}

\subsection*{Strengthened Inclusion Guarantees}

\begin{lemma}[$f$-Certified inclusion]
\label{lem:certified-inclusion}
Let $n\geq4f+1$, $q=n-f$, and $\FR=(q-2f)/(n-f)$.
The protocol satisfies the following strengthened pre-approval inclusion
property: if at most $f$ validators are Byzantine and a trade request $\tx$
obtains valid pre-approvals for an epoch from $q$ distinct validators, not
necessarily all honest, then $\tx$ is included in that epoch's final block.
\end{lemma}
\begin{proof}
At least $q-f$ approvers are honest. Since only $n-M_B$ validators lie
outside $V^*$, at least $q-f-(n-M_B)=M_B-2f$ honest approvers have selected
mini-blocks, each containing its approval of $\tx$. Moreover,
\[
 M_B-2f-\FR M_B
 =\frac{2f\bigl(M_B-(n-f)\bigr)}{n-f}\geq0.
\]
Thus $\tx$ appears in at least $\FR M_B$ selected approved sets and is
included by the pre-approved inclusion rule.
\qed
\end{proof}

\begin{lemma}[Strengthened accountable inclusion]
\label{lem:strong-accountable-inclusion}
Let $n\geq4f+1$, $q=n-f$, and $\FR=(q-2f)/(n-f)$.
There exists a publicly verifiable function $J$ such that, if a trade request
$\tx$ obtains an approval certificate but is omitted from a valid final block
for that epoch, there is evidence $w$ with $|J(w)|\geq f+1$ (and hence at
least $f$). Moreover, $v\notin J(w)$ for every honest validator $v$ and every
$w$.
\end{lemma}
\begin{proof}
Use the certificate and authenticated decided mini-blocks as evidence $w$.
Let $J(w)$ contain each selected certificate signer whose mini-block omits
its signed approval, and set $J(w)=\emptyset$ for invalid evidence.
At least $q+M_B-n=M_B-f$ certificate signers are selected. If $\tx$ is
omitted, fewer than $\FR M_B$ selected approved sets contain it. Therefore
\[
 |J(w)|>M_B-f-\FR M_B
 =\frac{2fM_B}{n-f}-f\geq f.
\]
Since $|J(w)|$ is an integer, at least $f+1$ validators are identified.
An honest validator includes every approval it issued for the epoch in its
mini-block, so valid evidence never identifies an honest validator.
\qed
\end{proof}

\end{document}